\documentclass{aa}  

\usepackage{graphicx}

\usepackage{txfonts}

\usepackage{url}
\usepackage{tabularx}
\usepackage{threeparttable}
\usepackage{xcolor}
\usepackage{multirow}

\usepackage{makecell}

\usepackage[colorlinks=true, linkcolor=blue, citecolor=blue, urlcolor=blue]{hyperref}

\begin{document}

   \title{Late-time X-ray afterglows of gamma-ray bursts: Implications for particle acceleration at relativistic shocks}

   \author{Zhi-Qiu Huang,
          \inst{1}
          Om Sharan Salafia,
          \inst{2,3}
          Lara Nava,
          \inst{2,4}
          Annalisa Celotti,
          \inst{1,2,4,5}
          \and
          Giancarlo Ghirlanda\inst{2}
          }

   \institute{SISSA, Via Bonomea 265, I-34136 Trieste, Italy
         \and
             INAF - Osservatorio Astronomico di Brera, Via E. Bianchi, 46, I-23807 Merate, Italy
             \and
             INFN - Sezione di Milano-Bicocca, Piazza della Scienza 3, I-20126 Milano, Italy
             \and
             INFN - Sezione di Trieste,Via Valerio 2, I-34127 Trieste, Italy
             \and
             IFPU - Institute for Fundamental Physics of the Universe, Via Beirut 2, I-34151 Trieste, Italy
             }

\date{Accepted XXX. Received YYY; in original form ZZZ}

  \abstract
   {Particle-in-cell (PIC) numerical simulations are currently among the most advanced tools for investigating particle acceleration at relativistic shocks. They are still limited by the finite computing power, whose effect is not straightforward to evaluate a priori. Observational features are hence required as verification.
   Gamma-ray burst (GRB) afterglow emission, produced by high-energy electrons accelerated at external shocks, provides a testbed for such predictions.
   Current numerical studies suggest that in GRB afterglows the maximum synchrotron photon energy, which corresponds to the limit of electron acceleration, may fall within the $\sim$ 0.1--10 keV X-ray energy band at late times, $t\gtrsim 10^6 - 10^7$ s. To test this prediction, we analyzed the X-ray spectra of six GRBs  with \emph{Swift}/XRT detections beyond $10^7$ s: our analysis revealed no clear evidence of a spectral cutoff. 
   Using a model that accounts for the effect of the finite opening angle of the shock on the observed maximum synchrotron photon energy, we show that these observations are incompatible with PIC simulation predictions, unless one or more physical afterglow parameters attain values that are at odds with those typically inferred from afterglow modeling (low radiative efficiency, low ambient density, and large equipartition fraction $\epsilon_{\rm B}$ of the magnetic field). 
   These findings challenge existing numerical simulation results and imply a more efficient acceleration of electrons to high energies than seen in PIC simulations, with important implications for our understanding of particle acceleration in relativistic shocks.}


   \titlerunning{Late-time X-ray detections of GRBs and particle acceleration}
   \authorrunning{Z.-Q. Huang et al. }

   \maketitle
\nolinenumbers

\section{Introduction}

The emission properties of gamma-ray bursts (GRBs) have been extensively studied through numerous observations. According to the standard model, an external relativistic shock forms and propagates outward through the surrounding medium. As the shock propagates, it interacts with the ambient material and accelerates electrons to very high energies \citep{1994ApJ...432..181M, 1998NewA....3..157D}. These accelerated particles produce a long-lasting afterglow emission through both the synchrotron and the synchrotron self-Compton (SSC) processes \citep{1998ApJ...497L..17S, 1999ApJ...512..699C, 2001ApJ...548..787S}, with detectable emission persisting in some cases for several months after the burst.

Numerical simulations suggest that weakly magnetized relativistic shocks that are characteristic of GRB afterglows are mediated by the Weibel instability \citep{1959PhRvL...2...83W, 1999ApJ...526..697M, 2005AIPC..801..345S, 2008ApJ...673L..39S, 2009ApJ...707L..92S, 2013ApJ...771...54S}. Particles undergo repeated scattering by small-scale turbulence generated through the instability, allowing them to cross the shock front multiple times and gain energy. The afterglow radiation produced by these accelerated electrons contains detailed information about particle acceleration at relativistic shocks.  

The maximum synchrotron photon energy, directly related to the maximum accelerated electron energy, is crucial for studying the detailed acceleration process as well as the microphysical conditions of the environment near the shock. An often invoked limit for the maximum energy (e.g., \citealt{2010ApJ...718L..63P, 2014Sci...343...42A, 2014A&A...564A..77P}) is the so-called synchrotron burnoff limit, where the synchrotron cooling rate is balanced by the idealized fastest Bohm-diffusion acceleration rate \citep{1996ApJ...457..253D, 2009herb.book.....D}, which can hardly occur. In numerical simulations by \citet{2013ApJ...771...54S}, where more realistic acceleration processes are considered, the acceleration rate is found to be slower and the predicted maximum energy lower than in the case of the synchrotron burnoff limit.

Starting from the result of \citet{2013ApJ...771...54S}, the maximum synchrotron photon energy emitted by accelerated electrons might move into the energy band of the X-Ray Telescope (XRT; 0.3-10 keV) on board the Niel Gerhels \textit{Swift} Observatory during the late-time phase ($10^6$ - $10^7$ s after the burst), producing observable features in the X--ray spectra. On June 11, 2025, the \textit{Swift} XRT catalog contained 1745 GRBs\footnote{\url{https://swift.gsfc.nasa.gov/archive/grb_table/stats/}} \citep{2007A&A...469..379E, 2009MNRAS.397.1177E, 2023MNRAS.518..174E}. We selected six events with XRT detections later than $10^7$ s and with a measured redshifts and analyzed their late-time X-ray spectra. By comparing the theoretical predictions with the results of our spectral analysis, implications can be obtained about the shock acceleration.

This paper is organized as follows. In section~\ref{sec:max_syn} we summarize the predictions of \citet{2013ApJ...771...54S} on the expected energy of synchrotron photons emitted by electrons reaching the acceleration limit during the late-time afterglow phase. We then refine these predictions by taking into account that the observed radiation originates from an 
equal-arrival-time surface (EATS). The procedure and results of the data analysis of the selected GRB afterglows are described in section~\ref{sec:data}. The constraints on the physical GRB parameters inferred from the comparison between the theoretical predictions and the observations are presented in section ~\ref{sec:constraints}. In section~\ref{sec:conclusion} we discuss the results and draw our conclusions.

\section{Maximum synchrotron photon energy}
\label{sec:max_syn}

\subsection{Analytical prediction based on PIC simulations}
\label{sec:max_hnu}

In this section, we recall the results by \cite{2013ApJ...771...54S} on the maximum electron energies under the assumption that the shock is ultra-relativistic. When the cooling rate of electrons balances the acceleration rate (we only considered the energy losses due to synchrotron emission), electrons can no longer gain energy and reach the  cooling limit at a Lorentz factor $\gamma_{\rm syn}$, corresponding to the typical synchrotron photon energy $h\nu_{\rm syn}$. At the acceleration rate estimated by \cite{2013ApJ...771...54S}, $h\nu_{\rm syn}$ can be expressed as
\begin{equation}
h\nu_{\rm  syn} \simeq 
 \left\lbrace
 \begin{array}{ll}
 620\ E_{\rm k , 54}^{1/4}\ n_{0}^{-1/12} \epsilon_{{\rm B},-3}^{-1/6}\, (1+z)^{-1/4} t_{{\rm obs},7}^{-3/4} \, \, {\rm keV},\, &\,   {\rm ISM}\\
 422\ E_{\rm k , 54}^{1/3}\ A_{\star}^{-1/6} \epsilon_{{\rm B},-3}^{-1/6}\ (1+z)^{-1/3}\, t_{{\rm obs},7}^{-2/3 \,}\, \, {\rm keV},\, &\,   {\rm wind}
 \end{array}
 \right.
 \label{eq:Sironi_syn},
\end{equation}
where $E_{\rm k}$ is the isotropic shock kinetic energy, $\epsilon_{\rm B}$ is the fraction of the shock energy that is carried by the downstream magnetic field, $t_{\rm obs}$ is the observer time, and $z$ is the redshift of the source. "ISM" and "wind" refer to the interstellar medium-like case (constant upstream number density $n = n_0$) and to the wind-like case ($n = A\, R^{-2}$, where $R$ is the distance from the progenitor, and $A_{\star} = A/3 \times 10^{35} {\, \rm cm^{-1}}$ \citealt{Chevalier2000}), respectively.  We use the notation $Q_{\alpha} = Q/10^{\alpha}$ for all the quantities in cgs units. The above estimate of $h\nu_{\rm  syn}$ neglects cooling induced by the synchrotron self-Compton (SSC) process, which is generally negligible at the times relevant to this study. This choice is conservative with respect to our conclusions. 

In addition, the acceleration process is quenched when the isotropization rate of downstream particles becomes lower than their gyro-frequency, in which case downstream particles are constrained to move along the magnetic lines perpendicular to the shock normal and can hardly cross the shock. This occurs when the particle energy reaches the so-called saturation, or magnetized, limit $\gamma_{\rm sat}$, corresponding to a characteristic synchrotron photon energy \citep{2013ApJ...771...54S}
\begin{equation}
h\nu_{\rm  sat} \simeq 
 \left\lbrace
 \begin{array}{ll}
2.85\ E_{\rm k , 54}^{1/2}\ \epsilon_{{\rm B},-3}^{1/2}\ \sigma_{\rm u,-9}^{-1/2}\ (1+z)^{1/2}\ t_{{\rm obs},7}^{-3/2} \, \, {\rm keV},\,\, &\,   {\rm ISM}\\
 2.10\ E_{\rm k , 54}^{1/2}\ \epsilon_{{\rm B},-3}^{1/2}\ \sigma_{\rm u,-9}^{-1/2}\  (1+z)^{1/2}\ t_{{\rm obs},7}^{-3/2 \,}\,\, {\rm keV},\, &\,   {\rm wind}
 \end{array}
 \right.
 \label{eq:Sironi_sat},
\end{equation}
where $\sigma_{\rm u}= \frac{B_{\rm u}^2}{4\pi n m_{\rm p}c^2}$ is the upstream magnetization parameter, $B_{\rm u}$ is the strength of the upstream magnetic field, and $m_{\rm p}$ is the proton mass. In the wind case, we assumed a toroidal magnetic field in the stellar wind that evolves as $B_{\rm u} \propto R^{-1}$. Hence, $\sigma_{\rm u}$ is a constant.

In the following, instead of using $\sigma_{\rm u}$, we explicitly specified the values of the density and upstream magnetic field, indicated as $B_{\rm ISM}$ and $B_{\rm W}$ for the ISM and wind case, respectively\footnote{Following \citet{2012ApJ...749...80S}, we considered $B_{\rm W}$ and $n$ at $R = 10^{18}$ cm for the wind case.}. We also define a quantity $\Psi = E_\mathrm{k, 54}\epsilon_\mathrm{B, -3}n_0/B^2_\mathrm{ISM, -6}$ ($\Psi = E_\mathrm{k, 54}\epsilon_\mathrm{B, -3}A_{\star}/B^2_\mathrm{W, -6}$) in the ISM (wind) case, so that $h\nu_{\rm sat} \propto \Psi^{1/2}$.

\begin{figure*}[t]
    \centering
    \includegraphics[width=0.45\linewidth]{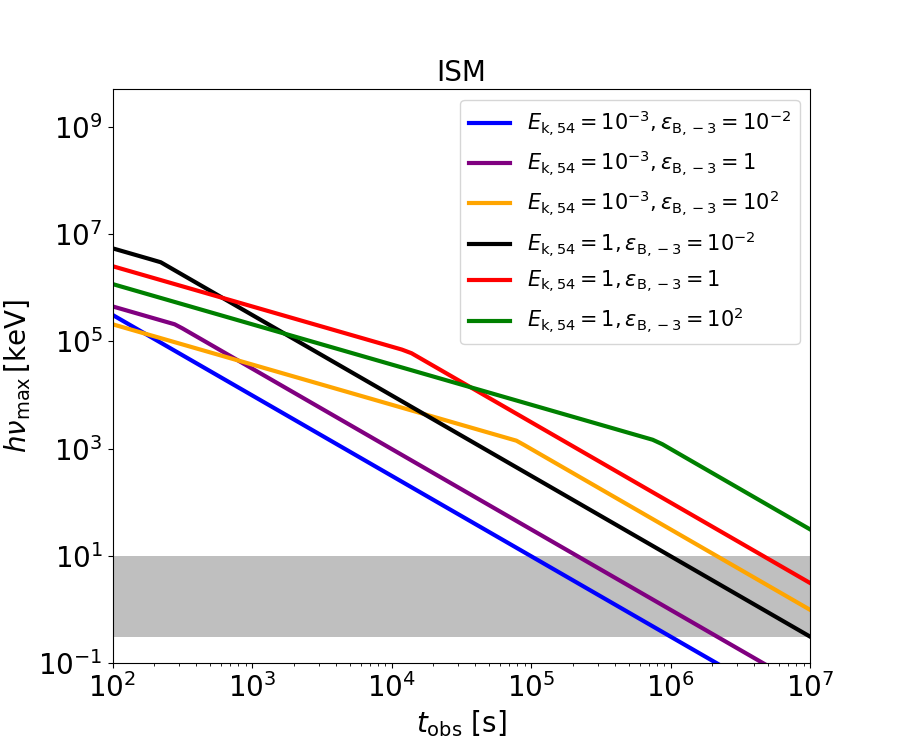}
    \includegraphics[width=0.45\linewidth]{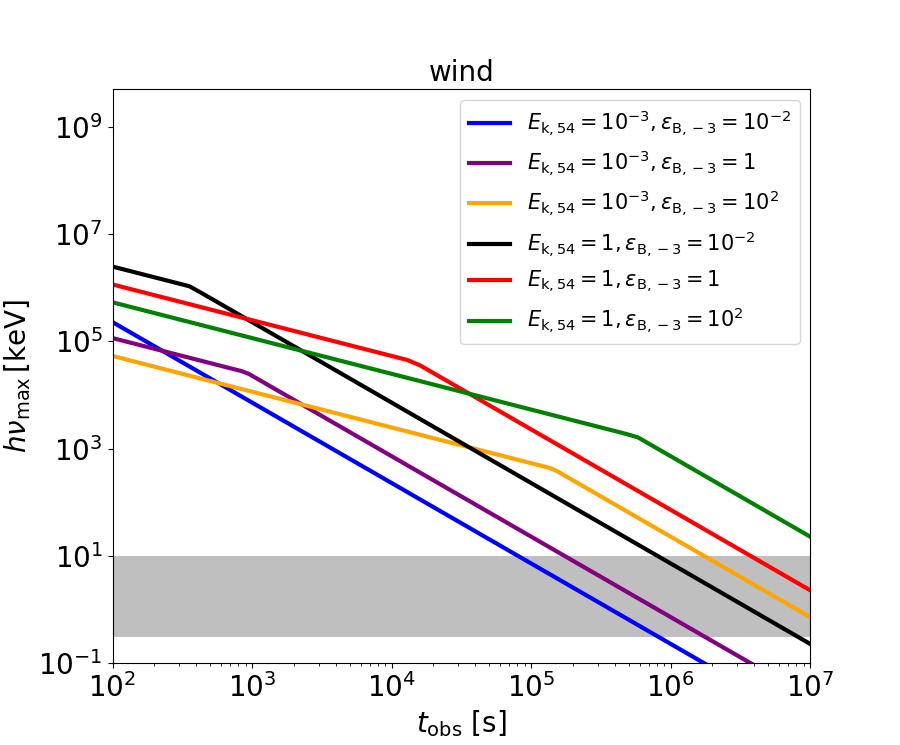}
    \caption{Time evolution of the maximum synchrotron photon energy $h\nu_{\rm max} = \min \, [h\nu_{\rm syn}, \, h\nu_{\rm sat}]$ for different sets of afterglow parameters. The redshift is fixed at $z=0.2$, and the upstream magnetization parameter is fixed to be $\sigma_{\rm u} = 10^{-9}$. The gray shaded area indicates the energy band between 0.3 and 10 keV.
    Left panel: ISM case with $n_0 = 0.1$. Right panel: Wind case with $A_{\star} = 1$. For $t_{\rm obs} \geqslant 10^6$ s, $h\nu_{\rm max}$ is set by $h\nu_{\rm sat}$ for all sets of parameters and can fall in the XRT energy band. \label{fig:maximum_energy}}
\end{figure*}

In Fig.~\ref{fig:maximum_energy} we plot the maximum attainable photon energy $h\nu_{\rm max} = \min \, [h\nu_{\rm syn},h\nu_{\rm sat}]$ for different sets of afterglow parameters. A redshift $z = 0.2$ was adopted as a benchmark, since the redshifts of the six GRBs that we selected for the  analysis lie between 0.07 and 0.35. For a typical range of values of the afterglow parameters, $h\nu_{\rm syn}$ and $h\nu_{\rm sat}$ intersect at $t_{\rm obs}$ much earlier than $10^6$ s, implying that for $t_{\rm obs} \geqslant 10^6$ s, $h\nu_{\rm max}$ is determined by $h\nu_{\rm sat}$. 

For GRBs with low energy and/or small $\epsilon_B$, $h\nu_{\rm max}$ falls within the 0.3 - 10 keV energy band (gray shaded area in Fig.~\ref{fig:maximum_energy}) at about $10^6$\,s. This would result in an observational feature (a cutoff in the XRT spectra) that can be searched for in late-time X-ray observational data.

\subsection{Effects of emission from high latitudes}
\label{sec:eats}

The predictions for the maximum synchrotron photon energy summarized in the previous section were derived for emission from material moving along the line of sight. For an extended source expanding radially at relativistic speed, photons that are received by the observer at the same time originate at different latitudes and at different times.  Due to the small radial width of the shock downstream, the region contributing the radiation observed at a fixed observer time approximately takes the form of a surface in space-time, the so-called equal-arrival-time surface (EATS; \citealt{1998ApJ...493L..31P, 1998ApJ...494L..49S, 2007ChJAA...7..397H}), where radiation from higher latitudes is emitted earlier than that from lower latitudes. This surface approximation can be adopted since synchrotron photons with the maximum energy are produced by electrons concentrated near the shock front. 

The evolution of the shock can be either adiabatic or radiative. In the radiative case, two conditions need to be satisfied. The fraction of the shock energy in relativistic electrons downstream $\epsilon_{\rm e}$ is expected to be close to 1 \citep{1998ApJ...497L..17S}, which contradicts the value of $\epsilon_{\rm e}$ from previous afterglow modeling $\epsilon_{\rm e} \lesssim 0.1$ (e.g., \citealt{2023MNRAS.518.1522D, 2023ApJ...956...12I, 2024MNRAS.527.6752G}). Moreover, the shock is expected to be in the fast-cooling case \citep{1998ApJ...497L..17S}. The transition from fast to slow cooling, however, occurs much earlier than the time in which we are interested \citep{2001ApJ...548..787S}. In the following, we therefore focus on the adiabatic case, and the relation between the shock Lorentz factor $\Gamma_{\rm sh}$ and its radius $R$ can be written as \citep{1976PhFl...19.1130B}
\begin{equation}
    \Gamma_{\rm sh}(R) = \left[ \frac{(17-4k)\ E_{\rm k}}{8\pi\ A\ m_{\rm p}c^{2}\ R^{3-k}}  \right]^{1/2},
    \label{eq:Gsh_of_R}
\end{equation}
for a relativistic blast wave produced by a point explosion, where $k = 0$ for the ISM case or $k=2$ for the wind case. The relation between the shock radius and the central engine frame time $t$ is
\begin{equation}
    R \approx c\ t/\left[1+ \frac{1}{2(4-k)\ \Gamma_{\rm sh}^2}\right].
    \label{eq:R_of_t}
\end{equation}

Photons emitted at a time $t$ by a portion of the shock located at a latitudinal angle $\theta$ (with respect to the line of sight) reach the observer at a time
\begin{equation}
    t_{\rm obs}/(1+z) = t-R \cos \theta /c.
    \label{eq:t}
\end{equation}

Combining the above equations, we obtain an implicit equation that links the shock radius, the observer time, and the latitude \citep{1997ApJ...489L..37S}, namely 
\begin{equation}
R(\theta, t_{\rm obs})=\frac{c\ t_{\rm obs}/(1+z)}{1-\cos \theta + 1/\left[2(4-k)\ \Gamma_{\rm sh}^2(R)\right]},
\label{eq:r}
\end{equation}
which can be solved numerically to determine the EATS, namely $R(\theta)$ for any fixed observer time. 

We now illustrate how the observed radiation, and in particular, the maximum observable synchrotron photon energy, is affected when the EATS is taken into account. In the shock downstream comoving frame, the synchrotron power emitted by an electron is $P^{\prime}_{\rm syn}(\gamma_{\rm e}) = \sigma_{\rm T} c\ \gamma_{\rm e}^2\ B^{\prime 2}/6\pi$, and the characteristic synchrotron frequency is $\nu^{\prime} (\gamma_{\rm e}) = \gamma_{\rm e}^{2} eB^\prime/2 \pi\ m_{\rm e} c$, where $\gamma_{\rm e}$ is the electron Lorentz factor, $\sigma_{\rm T}$ is the Thomson cross section, and $m_{\rm e}$ is the electron mass.
$B^{\prime}$ is the strength of the comoving magnetic field, which can be expressed as  
$B^\prime\, =\, (32 \pi\ m_{\rm{p}} c^2 \epsilon_{\rm{B}} n)^{1/2}\Gamma$.
The peak spectral power can be approximated as $P^{\prime}_{\nu, \rm max} \sim P^{\prime}_{\rm syn}(\gamma_{\rm e})/\nu^{\prime} (\gamma_{\rm e}) = \sigma_{\rm T}m_{\rm e}c^2 B^\prime/3e$.

From optically thin radiative transfer, the radiation intensity at the shock, as measured in the downstream comoving frame, is $I^{\prime}_{\nu^{\prime}} = j^{\prime}_{\nu^{\prime}}\Delta^{\prime}$, where $\Delta^{\prime}$ is the effective width of the downstream region, and $j^{\prime}_{\nu^{\prime}}$ is the synchrotron emissivity, given by
\begin{eqnarray}
 \label{eq:j}
    j^{\prime}_{\nu^{\prime}}(\nu^{\prime}) &&= \frac{1}{4\pi} \int^{+\infty}_{\gamma_{\rm min}}\frac{dn^{\prime}}{d\gamma_{\rm e}}(\gamma_{\rm e})\ P^{\prime}_{\nu^{\prime},\rm syn}(\gamma_{\rm e},\nu^{\prime})\ d\gamma_{\rm e} \\
    \nonumber
    &&\sim \frac{n^{\prime} P^{\prime}_{\nu, \rm max}}{4\pi}f(\nu^{\prime}).
\end{eqnarray}
Here, $f(\nu^{\prime})$ encodes the various power-law regimes that arise from the above convolution of the single-particle synchrotron specific power $P^\prime_{\nu^\prime,\rm syn}$ with the electron energy distribution $dn^\prime /d\gamma_{\rm e}$. 
The latter is the result of electron acceleration at the shock and the consequent radiative (and eventually adiabatic) cooling as electrons are advected in the downstream region. For typical GRB afterglow parameters, and focusing on late times as those of our interest, this results \citep{1998ApJ...497L..17S, 2001ApJ...548..787S, 2002ApJ...568..820G} in a broken power law, 
\begin{equation}
\frac{dn^\prime}{d\gamma_{\rm e}}\propto \left\lbrace\begin{array}{lr}
    {\gamma_{\rm e}}^{-p}  &\, \,  \gamma_{\rm min}<\gamma_{\rm e}\leq \gamma_{\rm c},  \\
     {\gamma_{\rm e}}^{-p-1} &\, \,  \gamma_{\rm c}<\gamma_{\rm e}< \gamma_{\rm max},\\
     0 & \rm otherwise,\\
\end{array}\right.
\end{equation}
where $p$ is the power-law index of the injected electron distribution, which extends between $\gamma_{\rm min}$ and $\gamma_{\rm max}$, and $\gamma_{\rm c}$ is the cooling Lorentz factor. 
Assuming $p\, >\, 2$ and $\gamma_{\rm{max}} \, \gg \, \gamma_{\rm{min}}$, we have 
\begin{equation}
\gamma_{\rm{min}} \, \sim \, \epsilon_{\rm{e}} \left( \frac{p-2}{p-1}\right) \frac{m_{\rm{p}}}{m_{\rm{e}}}\Gamma ,
\end{equation}
and $\Gamma = \Gamma_{\rm sh}/\sqrt{2}$ is the downstream Lorentz factor.
The cooling Lorentz factor can be estimated as 
$\gamma_{\rm c} m_{\rm e} c^{2} = P_{\rm syn}^{\prime}(\gamma_{\rm c})\, t^{\prime}$, where $t^{\prime} \simeq {R}/\Gamma c$. 

For $\gamma_{\rm max}$, we adopted the value estimated by \cite{2013ApJ...771...54S}, namely
\begin{equation}
    \gamma_{\rm max} = \frac{2\Gamma m_{\rm p}}{m_{\rm e}\sigma_{\rm u}^{1/4}}.
\end{equation}

The resulting spectral shape can thus be expressed as 
\begin{equation}
f(\nu^{\prime})=\left\{\begin{array}{ll}
\left(\frac{\nu^{\prime}}{\nu^{\prime}_{\rm{min}}}\right)^{1 / 3}  &\, \nu_{\rm{min}}^{\prime}>\nu^{\prime} \\
\left(\frac{\nu^{\prime}}{\nu^{\prime}_{\rm{min}}}\right)^{-(p-1) / 2} &\, \nu_{\rm{c}}^{\prime}>\nu^{\prime}>\nu_{\rm{min}}^{\prime} \\
\left(\frac{\nu^{\prime}_{\rm{c}}}{\nu^{\prime}_{\rm{min}}}\right)^{-(p-1) / 2}\left(\frac{\nu^{\prime}}{\nu^{\prime}_{\rm{c}}}\right)^{-p / 2}e^{-\nu^\prime/\nu^\prime_{\rm max}} &\, \nu^{\prime}>\nu_{\rm{c}}^{\prime}, 
\end{array}\right.
\end{equation}
where $\nu^\prime_{\rm min}$, $\nu^\prime_{\rm c}$, and $\nu^\prime_{\rm max}$ are the characteristic synchrotron frequencies of electrons with Lorentz factors $\gamma_{\rm min}$, $\gamma_{\rm c}$, and $\gamma_{\rm max}$, respectively. For simplicity, we assumed an exponential cutoff at $\nu^{\prime} = \nu^{\prime}_{\rm max}$ and neglected the synchrotron self-absorption, as it is not relevant at the frequencies of our interest.

Transformed to the observer frame, the radiation intensity is $I_{\nu}(\nu) = \mathcal{D}^3 I^{\prime}_{\nu^{\prime}}(\nu/\mathcal{D})= \mathcal{D}^3 j^{\prime}_{\nu^{\prime}}(\nu/\mathcal{D})\ \Delta^{\prime}$, where 
$\mathcal{D} = [{\Gamma (1- \beta \cos \theta )]^{-1}}$ is the relativistic Doppler factor. Since the downstream comoving electron density $n^{\prime} = N_{\rm e}/4\pi R^2 \Delta^{\prime}$, where $N_{\rm e}$ is the number of electrons swept by an isotropic spherical shock, we have (see Eq.~\ref{eq:j})   

\begin{equation}
    I_{\nu} = \mathcal{D}^3 \frac{N_{\rm e}\ P^{\prime}_{\nu, \rm max}}{16 \pi^2\  R^2}f(\nu/\mathcal{D}).
\end{equation}

For a source located at a luminosity distance $D_{\rm L}$,
the total observed flux density is obtained by integrating over the equal-arrival-time surface (the solution to Eq.~\ref{eq:r})
\begin{equation}
    F_{\nu}(t_{\rm obs}) \sim \frac{2\pi}{D_{\rm L}^2} \int R^2(\theta,t_{\rm obs})\ I_{\nu}(\nu / \mathcal{D},\theta,t_{\rm obs})\sin \theta\, d\theta,
\end{equation}
which leads to
\begin{equation}
    F_{\nu}(t_{\rm obs}) =\frac{1}{2}\int \mathcal{D}^2 F_{\nu, \rm max} \ f(\nu/\mathcal{D})\sin \theta \,d\theta,
\end{equation}
where we defined $F_{\nu, \rm max} \,  \equiv \, \mathcal{D} N_{\rm e} P^{\prime}_{\nu, \rm max}/4\pi D_{\rm L}^2$.

\begin{figure*}[t]
\begin{center}
\includegraphics[scale=0.4]{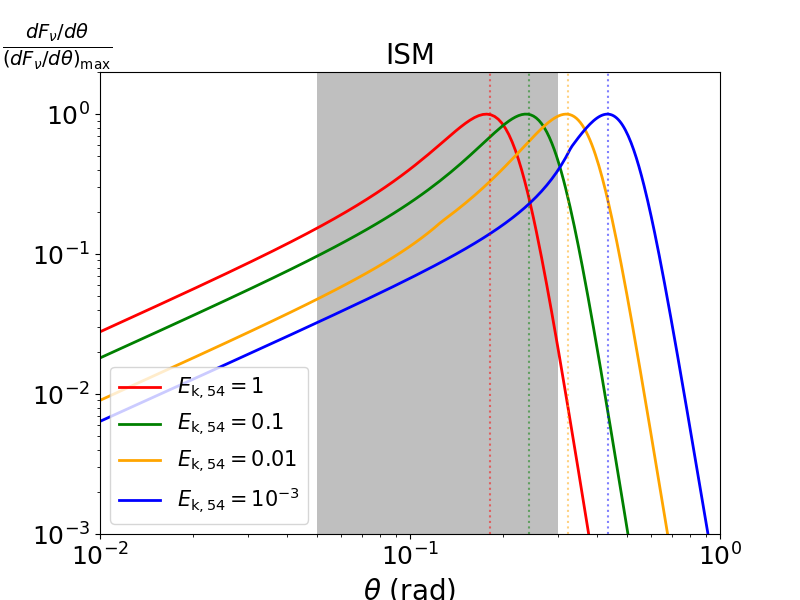}
\includegraphics[scale=0.4]{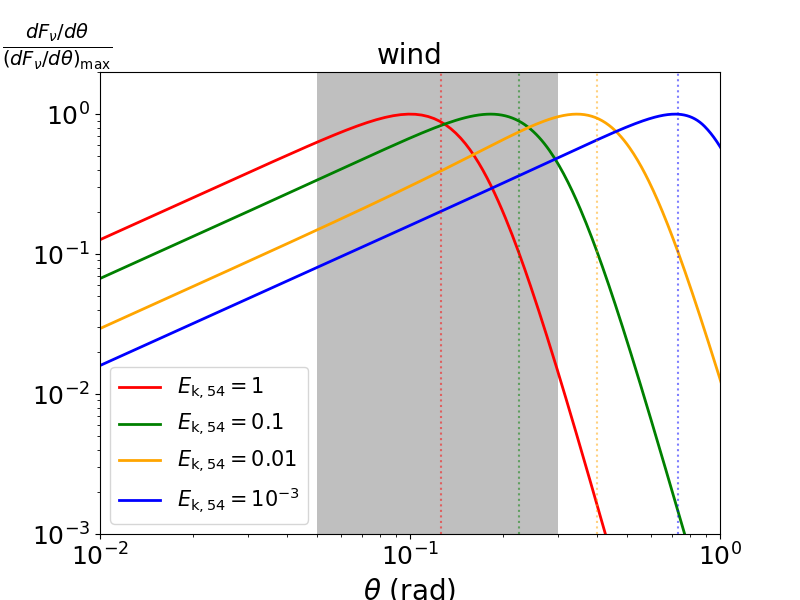}
\caption{Differential $dF_{\nu}/d\theta$ at $\nu/\mathcal{D} > \nu^{\prime}_{c}$
from different latitudes at the time $t = 3 \times 10^6$ s, adopting $p=2.2$ and $z=0.2$. The different line colors refer to different values of $E_{\rm k}$. The 
dashed vertical lines indicate the values of $\theta_{\rm c}$, and  
the gray shaded areas represent the typical range of values of the jet opening angle, $\theta_{\rm j} \in [0.05,0.3]$ radians. Left panel: ISM case with $n_{0} = 0.1$. Right panel: Wind case with $A_{\star} = 1$.\label{fig:flux1} }
\end{center}
\end{figure*}

The relative contribution to the flux from different latitudes can be seen
in Fig.~\ref{fig:flux1}, where we plot the differential flux $dF_{\nu}/d\theta$ at $\nu/\mathcal{D} > \nu^{\prime}_{c}$ as a function of the latitudinal angles (at the time $3 \times 10^6 s$).
This contribution increases with the latitudinal angle until it reaches a peak at a critical angle  $\theta_{\rm c}$, where $\theta_{\rm c}$ can be (numerically) estimated from
\begin{equation}
\theta_{\rm c} = \sin^{-1}\left[\frac{1}{\Gamma(R(\theta, t_{\rm obs}))}\right] \, 
\end{equation},
where $R(\theta, t_{\rm obs)}$ is given by Eq.~\ref{eq:r}. 

The critical angle $\theta_{\rm c}$ (marked with dotted vertical lines in Fig.~\ref{fig:flux1}) depends on the Lorentz factor of the shocked fluid. As time evolves, the shock decelerates and $\theta_{\rm c}$ increases. 
In the case of a jet geometry (expected to be the most likely for GRB ejecta), $\theta_{\rm c}$ is limited by the jet-opening angle $\theta_{\rm j}$. The gray area in Fig.~\ref{fig:flux1} represents a typical range for the jet-opening angle, $\theta_{\rm j} \in [0.05,0.3]$ radians \citep{2012ApJ...756..189F}. 
At the late times considered in this work, $\theta_{\rm c}$ can exceed $\theta_{\rm j}$. When this occurs, the 
flux is  mainly contributed by the edge of the jet rather than by $\theta_{\rm c}$.  
This can be seen in Fig.~\ref{fig:flux1}: when $\theta_{\rm j} < \theta_{\rm c}$, most of the high-energy flux originates at the jet edge, while  when $\theta_{\rm j} \gtrsim \theta_{\rm c}$, it 
predominantly arises from regions around $\theta_{\rm c}$. 

\begin{figure*}[t]
    \begin{center}
\includegraphics[scale=0.35]{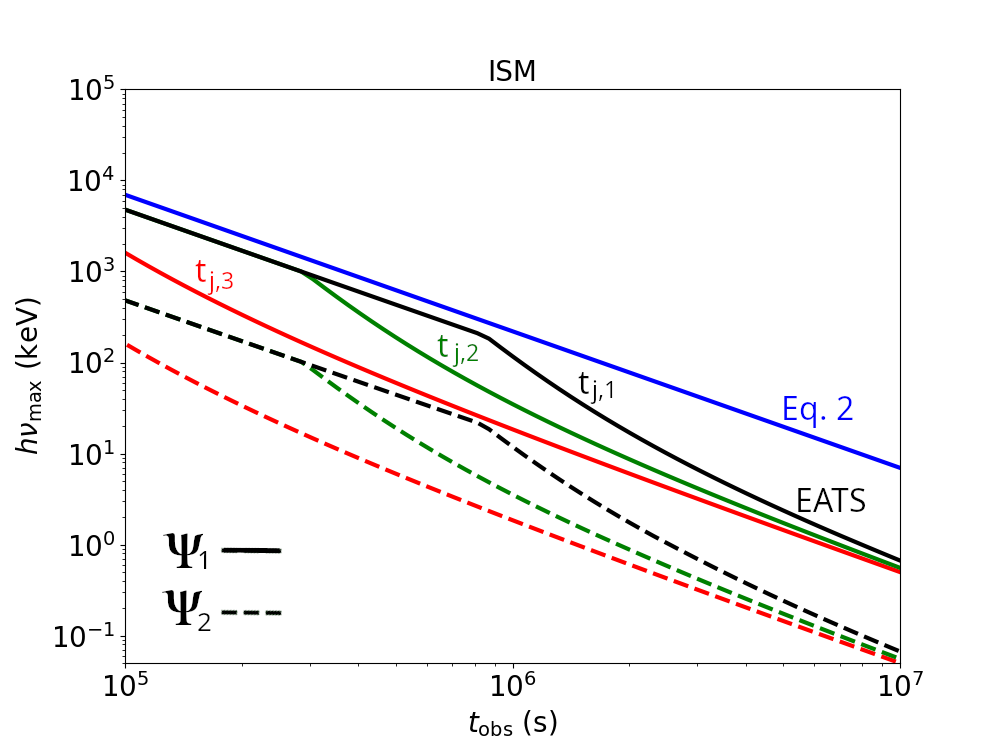}
\includegraphics[scale=0.35]{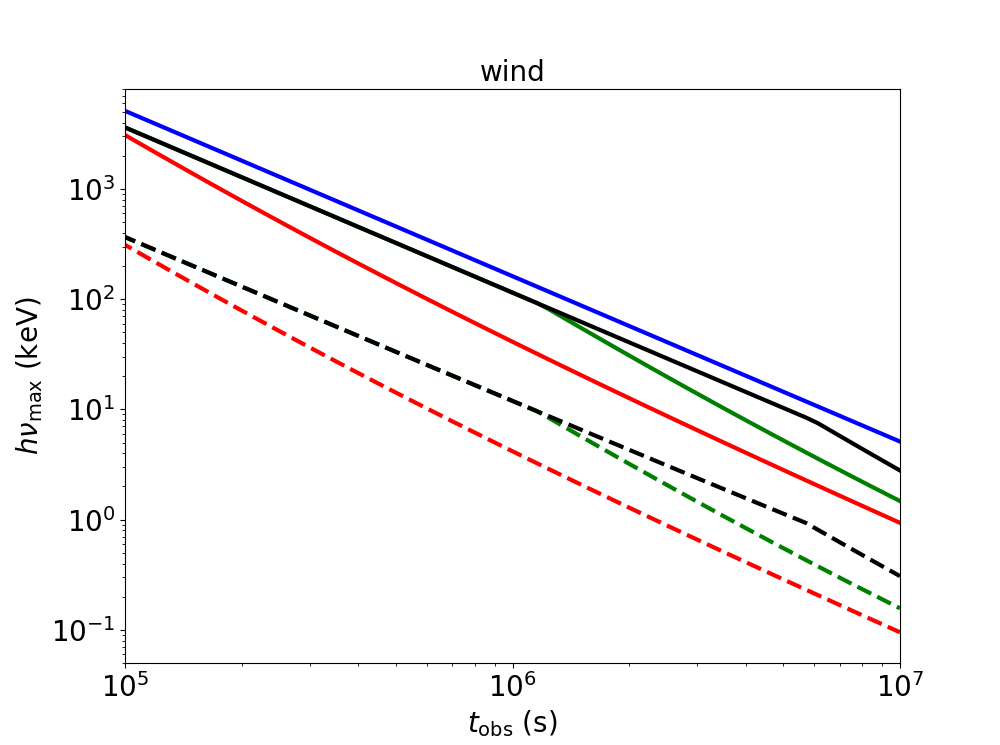}
\caption{Comparison between the maximum synchrotron photon energy predicted by Eq.\ \ref{eq:Sironi_sat} (blue curves, from \citealt{2013ApJ...771...54S}) and our prediction, obtained accounting for EATS and a conical geometry of the outflow (other curves). 
The black, green, and red curves correspond to three different jet break times $t_{\rm j}=t_{\rm j,1},$ $t_{\rm j,2}$, and $t_{\rm j,3}$. The solid curves are computed assuming the same value of $\Psi$ as the blue curve, $\Psi=\Psi_1$, and the dashed curves assume a lower $\Psi=\Psi_2=\Psi_1/100$. The left panel assumes a homogeneous ISM, and the right panel shows this for a wind-like external density profile.
\label{fig:compare} }
\end{center}
\end{figure*}

In order to assess the effect of the jet-opening angle on the observable $h\nu_{\rm max}$ as a function of $t_{\rm obs}$, we compare in Fig.\ \ref{fig:compare} the values of $h\nu_{\rm max}$ predicted by Eq.\ \ref{eq:Sironi_sat}, where the role of $\theta_{\rm j}$ is neglected (solid blue line), with those obtained when the emission is dominated by material at an angle $\min[\theta_{\rm c}, \theta_{\rm j}]$ (solid lines).

Each color corresponds to a different choice of jet break time $t_{\rm j}$ (as labeled), where  
\begin{equation}
\label{eq:tj_ISM}
\begin{split}
 & t_{\rm j} \sim (1+z)\, \theta_{\rm j}^{8/3} \left[\frac{17\ E_{\rm k}}{512\pi\ n\  m_{\rm p}c^5} \right]^{1/3}=  \\ & \quad = 1.4 \times 10^{5} (1+z) \left( \frac{\theta_{\rm j}}{0.1} \right)^{8/3} \left(\frac{E_{\rm k,54}}{n_0} \right)^{1/3} \, {\rm s}
\end{split}
\end{equation}
for the ISM case (left panel), and
\begin{equation}
\label{eq:tj_wind}
\begin{split}
    & t_{\rm j} \sim (1+z) \, \theta_{\rm j}^4 \left[\frac{9\ E_{\rm k}}{32\pi\ A\  m_{\rm p}c^3} \right] = \\
    & \quad  = 6.6 \times 10^{5} (1+z) \left( \frac{\theta_{\rm j}}{0.1} \right)^{4}\left(\frac{E_{\rm k,54}}{A_{\star}} \right) \, \, \, {\rm s}
    \end{split}
\end{equation}
for the wind case (right panel). 

The figure demonstrates that the impact of the shock anisotropy causes a deviation of $h\nu_\mathrm{max}$ from the prediction of Eq.\ \ref{eq:Sironi_sat} at times $t>t_\mathrm{j}$. This occurs because if the shock were spherical, the emission would be dominated by material at a latitude $\theta_\mathrm{c}>\theta_\mathrm{j}$. That material not being present, the actual value of $h\nu_\mathrm{max}$ is instead determined by the emission from $\theta_\mathrm{j}$.

In Fig.~\ref{fig:compare} we also compare the expected values of $h\nu_{\rm max}$ for parameters corresponding to the same jet break time $t_{\rm j}$ but different $\Psi$ (solid and dashed curves of the same color in the figure). 
While  $h\nu_{\rm max}$ are clearly different, their temporal dependence is not, being only determined by the jet break time  $t_{\rm j}$. As analytically shown in Appendix~\ref{appendix:B}, the 
normalized time evolution $h\nu_\mathrm{max}(t_\mathrm{obs})/h\nu_\mathrm{max}(t_\mathrm{j})$ only depends on the time variable $t_{\rm obs}/t_\mathrm{j}$. 
As a result, the evolution of $h\nu_\mathrm{max}$ is fully specified by $\Psi$ (which sets the normalization of Eq.\ \ref{eq:Sironi_sat})  
and by the normalized time variable $t_\mathrm{obs}/t_\mathrm{j}$.

\begin{figure*}[t]
    \begin{center}
\includegraphics[scale=0.35]{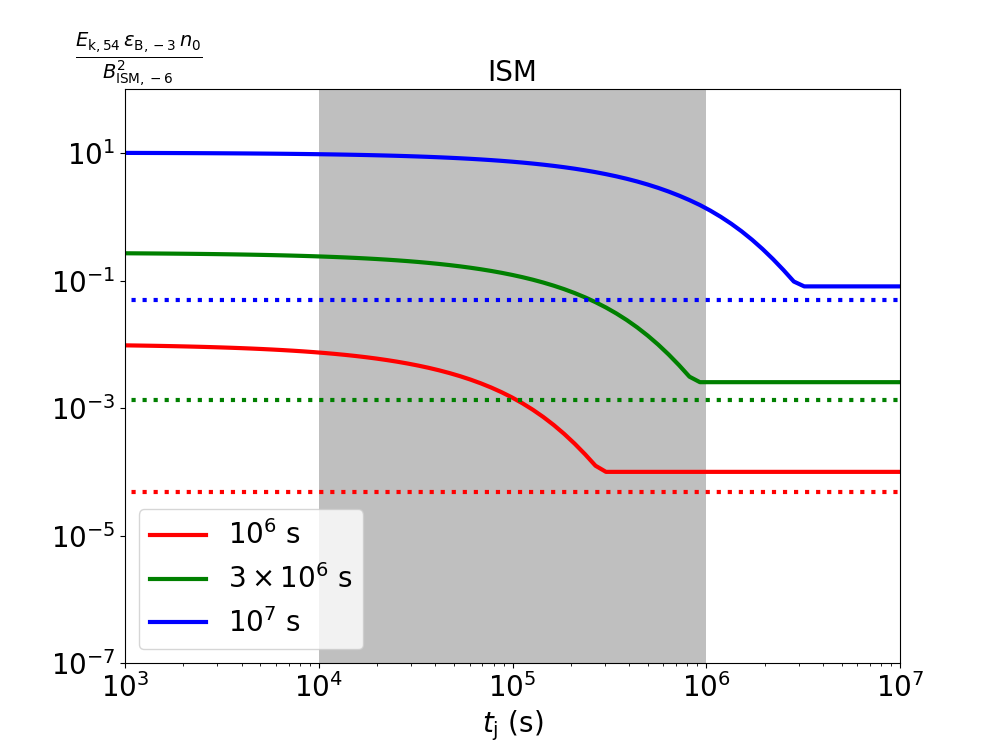}
\includegraphics[scale=0.35]{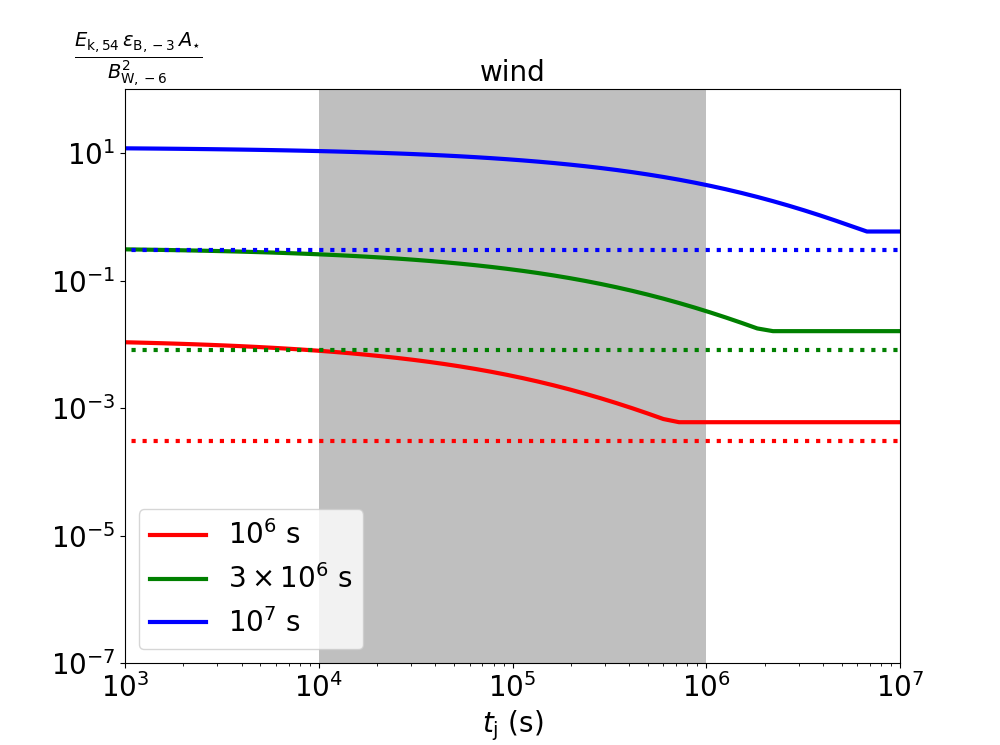}
\caption{Minimum allowed value of the parameter $\Psi = E_\mathrm{k, 54}\epsilon_\mathrm{B, -3}n_0/B^2_\mathrm{ISM, -6}$ (ISM, left panel) or $E_\mathrm{k, 54}\epsilon_\mathrm{B, -3}A_{\star}/B^2_\mathrm{W, -6}$ (wind, right panel) as a function of the assumed jet break time $t_{\rm j}$ for a synchrotron spectrum with maximum energy $h\nu_{\rm max} > 3$ keV observed at  $10^6$, $3 \times 10^6$ and $10^7$ s (for a source redshift $z=0.2$). The gray shaded areas indicate the range of typical values of jet break times, $t_{\rm j} \in [10^4,10^6]$ s. The solid lines show the results in this work, which account for the effect of EATS, and the dotted lines reproduce the predictions of \citet{2013ApJ...771...54S}. 
\label{fig:parameter} }
\end{center}
\end{figure*}

The jet break time $t_{\rm j}$, in turn, only depends on $\theta_{\rm j}$ and $E_{\rm k}/n$ (or $E_{\rm k}/A$; Eqs. ~\ref{eq:tj_ISM} and ~\ref{eq:tj_wind}). Therefore, as illustrated in Fig.~\ref{fig:compare}, a lower limit on $h\nu_{\rm max}$ at a given $t_{\rm obs}$ translates into a lower limit on $\Psi$ that depends on the choice of $t_{\rm j}$. Fig.~\ref{fig:parameter} shows example limits on $\Psi$ as a function of $t_{\rm j}$ for a hypothetical limit $h\nu_{\rm max} > 3$ keV at three different observed times.

The gray shaded area represents the range of typical values of the jet break time, $t_{\rm j} \in [ 10^4, 10^6]$ s, as inferred from observations \citep{2001ApJ...562L..55F, 2003ApJ...594..674B, 2004ApJ...616..331G, 2009ApJ...698...43R, 2010ApJ...711..641C}. The change in behavior in each (solid) curve  marks the transition between $\theta_{\rm c}$ and $\theta_{\rm j}$. The dashed lines represent the results from Eq.~\ref{eq:Sironi_sat}, which are consistent with our calculations when $\theta_{\rm j} > \theta_{\rm c}$.

\section{X-ray data analysis and results}
\label{sec:data}

To search for the presence of a cutoff, we selected events in the \textit{Swift} XRT sample (up to June 11, 2025) that satisfied the following criteria:
(i) number of counts accumulated between $10^6$ and $10^7$ s after the BAT trigger greater than $10^4$, to allow for a robust spectral analysis;
(ii) spectroscopically measured redshift;
(iii) measured isotropic-equivalent gamma-ray energy $E_\gamma$.

The six selected bursts were GRB 130427A, GRB 130702A, GRB 130925A, GRB 161219B, GRB 190829A, and GRB 221009A. Their XRT light curves, which were taken directly from the \emph{Swift}/XRT repository, are shown in Fig.~\ref{fig:index}. The redshifts and prompt emitted energies are reported in Cols. 2 and 3 of Tab.~\ref{tab:E_gamma}. As expected, our criteria selected events with a relatively low redshift ($z<0.35$), for which the probability of a still detectable X-ray flux at late times is higher. In terms of energetics, the sample spans a wide range of values that extend over four orders of magnitude. 

\begin{table}[t]
\centering
\caption{Gamma-ray bursts analyzed in this work, listed together with their redshift $z$ and the energy $E_{\gamma}$ that is emitted during the prompt phase.
\label{tab:E_gamma}}
\begin{tabular}{cccc}
\hline
Event       & $z$      & $E_{\gamma}$ (erg) & References  \\ \hline
GRB 221009A & 0.151    & $3 \times 10^{54}$  &  1, 2  \\
GRB 190829A & 0.0785  & $2 \times 10^{50}$ &  3, 4 \\
GRB 161219B & 0.1475  & $1.6 \times 10^{50}$ & 5, 6 \\
GRB 130925A & 0.347   & $1.5 \times 10^{53}$  & 7, 8\\
GRB 130702A & 0.145  & $6.4 \times 10^{50}$ & 9, 10\\
GRB 130427A & 0.34  & $8.5 \times 10^{53}$ & 11, 12 \\ \hline
\end{tabular}
\tablefoot{(1)~\cite{2022GCN.32648....1D}; (2)~\cite{2022GCN.32668....1F}; (3)~\cite{2019GCN.25565....1V}; (4)~\cite{2019GCN.25660....1T}; (5)~\cite{2016GCN.20321....1T}; (6)~\cite{2016GCN.20323....1F}; (7)~\cite{2013GCN.15249....1V}; (8)~\cite{2013GCN.15260....1G}; (9)~\cite{2013GCN.14983....1L}; (10)~\cite{2013GCN.14986....1G}; (11)~\cite{2013GCN.14455....1L}; (12)~\cite{2013GCN.14487....1G}. For each GRB, the first reference is for redshift, and the second reference is for $E_{\gamma}$.}
\end{table}

We downloaded their XRT spectral data at epochs $>10^5$ s from the \emph{Swift}/XRT repository\footnote{\url{https://www.swift.ac.uk/index.php}}  \citep{2007A&A...469..379E, 2009MNRAS.397.1177E, 2023MNRAS.518..174E}. For each GRB, the time intervals within which time-resolved spectra were extracted are reported in the third column of Tab.~\ref{tab:mcmc} and are highlighted in the corresponding light curves in Fig.~\ref{fig:index}) with different colors. 

The spectral analysis was performed with \texttt{Xspec v.12.15.0}. We tested two models: an absorbed power--law model, and an absorbed power law with a high-energy exponential cutoff. In both cases, we fixed the Galactic hydrogen--equivalent column density $N_{\rm H, Gal}$ to the value corresponding to the GRB coordinates \citep{1990ARA&A..28..215D, 2005A&A...440..775K, 2016A&A...594A.116H} as provided by the HEASARC service\footnote{\url{https://heasarc.gsfc.nasa.gov/cgi-bin/Tools/w3nh/w3nh.pl}} (see Col. 4 of Tab.~\ref{tab:E_gamma}). The intrinsic absorption $N_{\rm H}$ was left free to vary in the fits, but was set as a common parameter for the spectra corresponding to different time bins of the same GRB. 

We used the multiplicative model function $\mathtt{cflux}$, which returns the integrated deabsorbed fluxes $\log_{10}{\rm Flux \, (erg \, cm^{-2} \, s^{-1})}$ within a given energy band ($E_{\rm min} \, = \, 0.3$ keV and $E_{\rm max} \, = \, 10$ keV in our case) for the given spectral model. For the power-law model, the photon indices $\Gamma_{\rm XRT}$ ($dN/dE\propto E^{-\Gamma_{\rm XRT}}$) were left free to vary in the different temporal bins of the same GRB. When $h\nu_{\rm max}$ moved into the XRT band, a softening of the spectral index is expected. The photon indices obtained by the power-law fits are shown in Fig.~\ref{fig:index} in the panels below the corresponding light curves, where the time intervals selected for the spectral analysis are highlighted with different colors. According to our data analysis, the photon indices at different time intervals of the same GRB are consistent with each other within the error bars in 90$\%$ confidence ranges, and we found no clear evidence of a temporal softening, as expected when a spectral break were to cross the XRT energy band from higher to lower energies. 

For the model of a power law with an exponential cutoff, given the presence of an additional free parameter, we assumed (as supported by the results of the simple power-law model) that the photon index was independent of time for the spectra of the same burst (i.e.,\ we took it as a single free parameter common to all the time bins of the same burst), while the cutoff energy was left free to vary in different time bins. All fits were performed by exploring the free parameters with a Monte Carlo Markov chain (MCMC), and their values are listed in Tab.~\ref{tab:mcmc}. The values of $N_{\rm H}$ in Col. 5 were adopted from the best-fitting results using the power-law model. Only one-side lower limits on the cutoff energy were found: their values (at a confidence level of 2$\sigma$) are listed in the last column of the table.

As a comparison, we also fit the spectra with a single power-law model assuming a time-independent spectral index. Tab.~\ref{tab:C-stat} lists the corresponding C-statistic values \citep{1979ApJ...228..939C} and the differences in the Bayesian information criterion $\Delta$BIC between the two models for each GRB. Although the cutoff model provides slightly lower C-statistic values in most cases, the corresponding $\Delta$BIC values indicate that the improvement is not statistically significant given the additional free parameters. We therefore found no evidence of a spectral cutoff in the data.

\begin{table}[t]
\centering
\caption{Comparison of the modeling between the single power-law model with a time-independent spectral index (PL) and the power-law model with an exponential cutoff (cutoff) for six GRB events.
\label{tab:C-stat}}
\begin{tabular}{cccc}
\hline
Event       & C-stat (PL)      & C-stat (cutoff)  & $\Delta$BIC  \\ \hline
GRB 221009A & 451.37    & 449.17 & 13.28\\
GRB 190829A & 143.08  & 142.20 &  13.95\\
GRB 161219B & 251.20  & 243.89 & 16.23\\
GRB 130925A & 337.58   & 336.79  &  9.38\\
GRB 130702A & 346.92  & 343.63 & 8.67\\
GRB 130427A & 499.00  & 499.69  & 16.37\\ \hline
\end{tabular}
\tablefoot{
The C-statistic values of PL and cutoff models are listed in Cols. 2 and 3, respectively, and the differences in $\Delta{\rm BIC}= {\rm BIC}_{\rm cutoff} - {\rm BIC}_{\rm PL}$ are listed in the last column. Positive values of $\Delta{\rm BIC}$ indicate a preference for the PL model.
}
\end{table}

To constrain the physical parameters, information on the spectral cutoff energy is required. Lacking the cutoff energy from XRT data, we conservatively adopted the 2$\sigma$ lower limits from the cutoff model in our following calculations.

\section{Constraints from the absence of observed cutoffs}
\label{sec:constraints}

The lack of evidence of a spectral cutoff for the six GRBs we examined imposes a lower limit on $h\nu_{\rm max}$ at the time $t_{\rm obs}$ when it was inferred for each GRB. 
This in turn sets constraints on its physical parameters, as discussed in section \ref{sec:eats} (see Fig.~\ref{fig:parameter}). 
As explained in the previous sections, these constraints depend on the jet break time we assumed. The post-jet break light curve is expected to decay as $t^{-\alpha}$, with $3p/4\leq\alpha\leq p$ \citep{Gao2013,Rhoads1999}, depending on the density profile of the external medium (either uniform or wind-like) and on whether the forward shock significantly expands laterally. When the X-ray band is above $\nu_\mathrm{c}$, then the X-ray photon index $\Gamma_\mathrm{XRT}$ and the electron energy distribution index $p$ are related by $p = 2\Gamma_\mathrm{XRT}-2$. With these notions in mind, a simple visual inspection of Figure \ref{fig:index} shows that the late-time decay indices of the X-ray light curves of the examined GRB afterglows are all compatible with a post-jet-break phase, even though a pre-jet break can still be accommodated if $\nu_\mathrm{c}$ is instead above the X-ray band. In two cases (GRB 130925A and GRB 190829A), there is some evidence of a steepening at around $10^6$ seconds, but it is not possible to conclusively determine whether this is related to a jet break without an in-depth analysis of the multiwavelength dataset for each individual GRB. For these reasons, and for simplicity, we preferred to remain agnostic about the jet break time in our analysis and instead show our results as a function of $t_\mathrm{j}$.

We note that $E_{\rm k}$ is not directly observable, whereas the isotropic equivalent gamma-ray energy detected during the prompt phase $E_{\gamma}$ can be measured from observations. For a better connection between our constraints and the observed properties of the bursts, we defined the radiative efficiency $\eta_{\gamma}$ as
\begin{equation}
    \eta_{\gamma} = \frac{E_{\gamma}}{E_{\rm k}},
\end{equation}
where $E_{\gamma}$ is known for the six GRBs we studied (see Tab.~\ref{tab:E_gamma}).
This definition differs from that more commonly used $\eta_{\gamma}=\frac{E_{\gamma}}{E_{\gamma}+E_{\rm k}}$ \citep{2004ApJ...613..477L, 2007ApJ...655..989Z}: clearly, the two definitions are equivalent 
when $E_{\gamma} \ll E_{\rm k}$.
Estimates of the radiative efficiency have been reported in previous works \citep{2007ApJ...655..989Z, 2015ApJS..219....9W}: its typical value is $\eta_{\gamma}\sim 10 \%$ \citep{2016MNRAS.462.2990D, 2014MNRAS.442...20H, 2018ApJS..236...26L, 2012MNRAS.425..506D}, but it might be significantly higher in some GRBs \citep{2004ApJ...613..477L, 2024ApJ...972..195L}.

Using the above definition of radiative efficiency and leveraging the measurement of $E_\gamma$ available for our GRBs, we turned our lower limits on the cutoff energy at a given time into lower limits on the quantity $\Psi/E_{\gamma,53}=\epsilon_{\rm B, -3}n_0/\eta_{\gamma, -1}B^2_{\rm ISM, -6}$ (in the ISM case; $\Psi/E_{\gamma,53}=\epsilon_{\rm B, -3}A_{\star}/\eta_{\gamma, -1}B^2_{\rm W, -6}$ in the wind case) for each event in the sample. Specifically, for each selected burst, we set $t_{\rm obs}$ equal to the start time of the latest available XRT spectrum; for the lower limit on the cutoff energy, we adopted the 2$\sigma$ lower bound on $E_{\rm cut}$ inferred from the spectral analysis (as reported in Tab.~\ref{tab:mcmc}). These choices ensured that our derived constraints on the parameter space are conservative.

\begin{figure*}[t]
    \begin{center}
\includegraphics[scale=0.35]{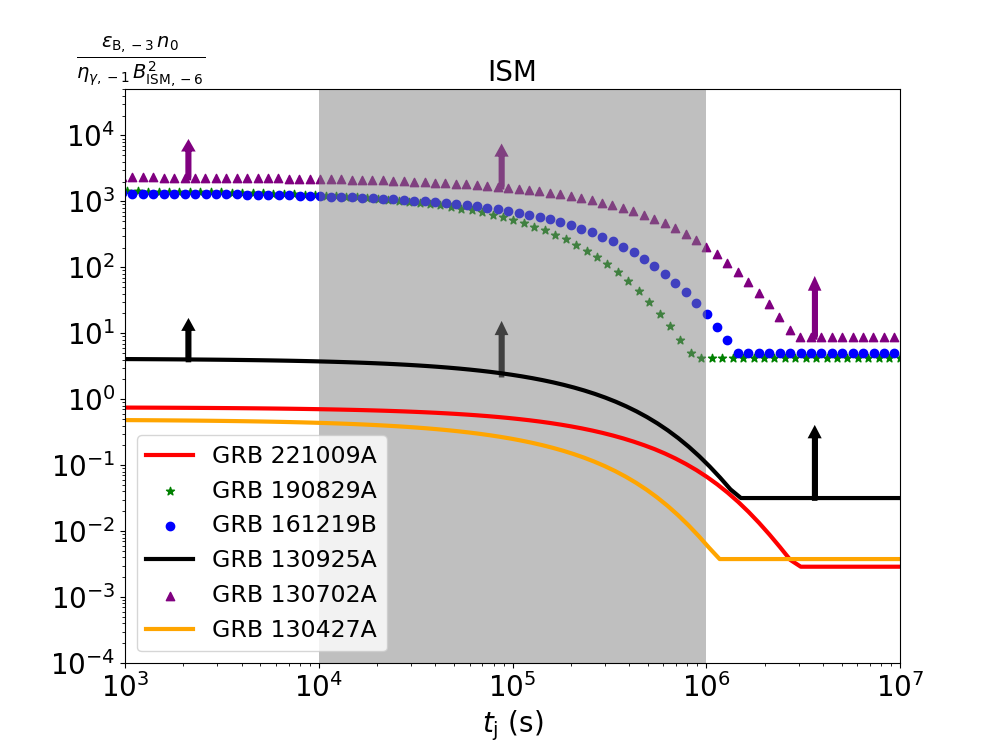}
\includegraphics[scale=0.35]{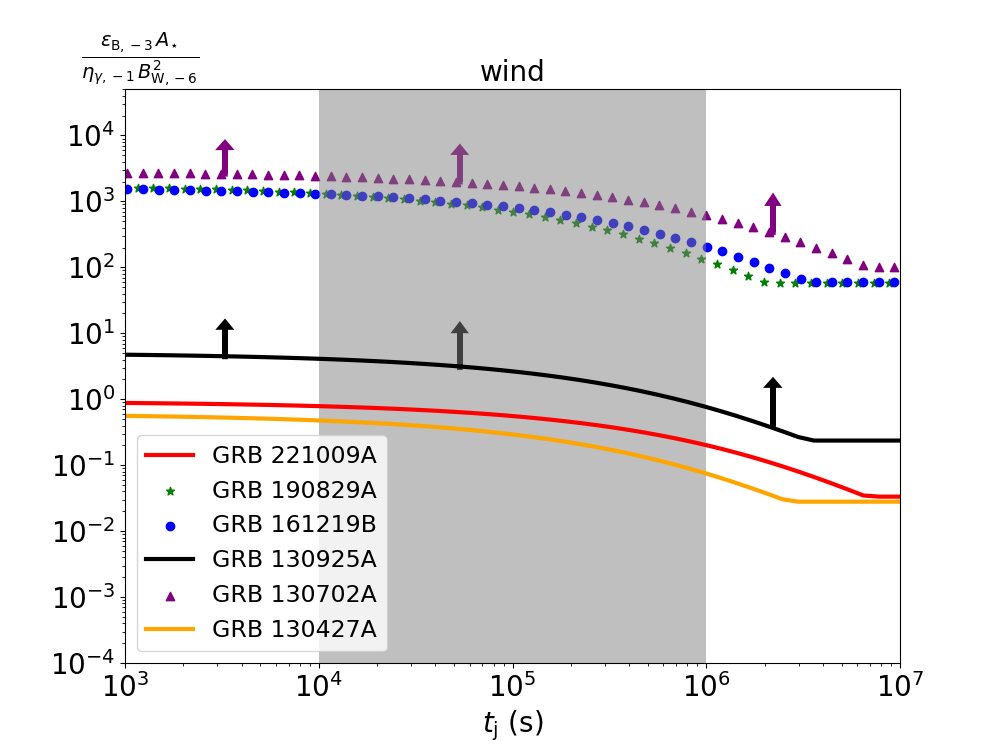}
\caption{Constraints on the parameter space derived from the spectral analysis of the six selected GRBs. Similar as Fig. ~\ref{fig:parameter}, but the quantity in the y-axis is defined as $\Psi/E_{\gamma,53}=\epsilon_{\rm B, -3}n_0/\eta_{\gamma, -1}B^2_{\rm ISM, -6}$ (ISM) or $\epsilon_{\rm B, -3}A_{\star}/\eta_{\gamma, -1}B^2_{\rm W, -6}$ (wind). Each curve was built considering the prompt emitted energy $E_{\gamma}$ of the GRB (listed in Tab.~\ref{tab:E_gamma}) and the lower limit on $h\nu_{\rm max}$ obtained from our spectral analysis at a given observational time $t_{\rm obs}$.
The gray shaded regions represent the typical range of the jet break time $t_{\rm j} \in [10^4, 10^6]$ s, and the values of the $y$-axis correspond to representative values of the relevant parameters.
The arrows indicate that the curves represent lower bounds on the allowed parameter space. 
Left panel: ISM case. Right panel: Wind case. The three GRBs with low $E_{\gamma}$, which differ for standard values of the parameters from the predictions on $h\nu_{\rm max}$, are marked with distinct symbols: green stars for GRB 190829A, blue circles for GRB 161219B, and purple triangles for GRB 130702A. 
\label{fig:data} }
\end{center}
\end{figure*}

For each GRB, the strongest of the derived lower limits is shown in Fig.~\ref{fig:data}. 
The gray shaded area indicates the range of typical values of the jet break time $t_{\rm j}$. In the figure, a vertical coordinate  $\Psi/E_\mathrm{\gamma,53}$ equal to unity would correspond to representative values of the relevant parameters, that is, $\epsilon_\mathrm{B}=10^{-3}$, $n=1\,\mathrm{cm^{-3}}$ (or $A_\star=1$), $B_\mathrm{ISM}=10^{-6}\,\mathrm{G}$ (or $B_\mathrm{W}=10^{-6}\,\mathrm{G}$) and $\eta_\gamma=10\%$.
The vertical arrows indicate that the parameter space consistent with the observations, specifically, the lower limit on the spectral cutoff energy
for each GRB is that above the corresponding curve. 

Since only lower limits on the cutoff energy could be derived, the resulting constraints for all six GRBs are marginally consistent with typical afterglow parameters at best unless their jet-break times are all rather late ($\gtrsim 10^6$ s).

The constraints are particularly tight, 
with $\Psi/E_\mathrm{\gamma,53}\gtrsim 10^{2}-10^3$,
for three GRBs, namely GRB 130702A, GRB 161219B, and GRB 190829A (their curves are marked with particular patterns): to account for the lack of a spectral cutoff in their spectra, we needed to invoke a very 
low radiative efficiency, a very large $\epsilon_{\rm B}$, or a very high external density (or some combinations of these factors).

These GRBs have particularly low prompt energetics, lower than $ 10^{51}$ erg \citep{2013GCN.14986....1G, 2016GCN.20323....1F, 2019GCN.25575....1L, 2019GCN.25660....1T}: since $h\nu_{\rm sat}\propto E_{\rm k}^{1/2}$, their late X-ray emission indeed provides the best opportunity to test for a cutoff or place tighter constraints on the parameter space based on its absence. 
We note that the $\gamma_\mathrm
{sat}$ model that we employed is based on PIC simulations of relativistic shocks with a Lorentz factor $\Gamma=15$, and it is expected to be valid for shocks that are at least mildly relativistic. According to the analysis of \citet{2013ApJ...771...54S}, however, the particle acceleration in mildly relativistic shocks is less efficient because the Weibel and oblique instabilities are suppressed. These instabilities generate the magnetic turbulence that governs the Fermi process. The shock in the three weakest events is expected to decelerate to Lorentz factors $\Gamma_{\rm sh} \lesssim 1.5$ unless the shock is particularly energetic or the external density is sufficiently low. In the ISM case, this quantitatively translates into having $\eta_{\gamma,-1}\, n_{\rm 0} \lesssim 0.05-0.5$.
For these three events, our results are therefore conservative with respect to our conclusions.

\section{Discussion and conclusions}
\label{sec:conclusion}
We analyzed the X-ray synchrotron spectra of six GRBs during their late-time afterglow phase to search for spectral cutoff signatures to probe the maximum  energy achievable by electrons accelerated at relativistic shocks.
The motivation was provided by the results of numerical simulations by \cite{2013ApJ...771...54S}, who predicted the expected maximum energy of electrons accelerated in relativistic weakly magnetized shocks, and consequently, the maximum achievable synchrotron photons energy.

We first  showed that these predictions would imply a spectral cutoff that might enter the XRT energy band at late times ($10^6-10^7$\,s). Before testing these predictions against observations, we refined the estimates on the maximum photon energy by accounting for EATS and the shock anisotropy induced by a conical jet geometry.
Our analysis focused on six GRB events with exceptionally late-time XRT detections. The data revealed no statistically significant evidence of a spectral cutoff within the XRT energy range. For each GRB, the 2$\sigma$ lower limit on the cutoff energy provided constraints on the afterglow parameters under the assumption that the numerical results of \citet{2013ApJ...771...54S} apply. We presented our results for an ISM and a wind environment. Although some (relatively small) quantitative differences are apparent between the two assumed environments, the overall conclusions of this work do not differ qualitatively between the two cases.

Fig.~\ref{fig:maximum_energy} shows that early-time (10--100\,s) observations at gigaelectronvolt energies can also provide valuable constraints on the maximum energy. Afterglow photons with energies above the gigaelectronvolt range have been detected in several events by the Large Area Telescope (LAT) on board the Fermi Gamma-ray Space Telescope, such as GRB 130427A and GRB 221009A (which are in our sample), and GRB 190114C \citep{2019Natur.575..459M}. Unfortunately, the low photon number statistics of the GRB 130427A and GRB 190114C observations makes a detailed reconstruction of the spectral shape challenging. This limits the possibility that valuable information on $h\nu_\mathrm{max}$ can be extracted from these datasets \citep{Wu2026}. For GRB 221009A, the LAT data do not reveal a clear cutoff either, possibly because this is masked by a strong SSC component \citep{Banerjee2025}.

For typical values of the 
afterglow parameters, 
the nondetection of a spectral cutoff in the X-ray band for the three most energetic GRBs in our sample is marginally
consistent at best with theoretical predictions.
The lower limit 
in the parameter space for these bursts (solid lines in  Fig.~\ref{fig:data}) clusters around values on the order of unity for the combination of parameters shown on the $y$-axis of the plot. This corresponds to typical afterglow  parameters, namely a magnetic field fraction $\epsilon_{\rm B}=10^{-3}$, a circumburst density $\sim 1$ cm$^{-3}$ (or $\sim A_\star$), and a prompt emission efficiency $\eta_\gamma=10\%$, for jet break times $t_{\rm j} \lesssim$ a few $10^5$ s. We note that afterglow modeling (e.g., \citealt{BeniaminiNava15,2019ApJ...884..117W, 2022ApJ...931L..19S, 2022ApJ...925..182H, 2023ApJ...946L..27A}) commonly indicates values of $\epsilon_{\rm B} $ 
in the range $10^{-5}$ to $10^{-3}$, which constrains the parameter space even more. 

For the three remaining GRBs (GRB 130702A, GRB 161219B, and GRB 190829A), which are significantly less energetic, the constraints on the parameter space from the late-time spectral analysis (symbol lines in Fig.~\ref{fig:data}) can only be satisfied by invoking a very low radiative efficiency, $\eta_{\gamma}<1\%$, or very large $\epsilon_{\rm B}\sim 1$ or densities ($ n \sim 10^3$ cm$^{-3}$).   
Low values of the prompt efficiency would imply an extremely strong requirement on the total energetics,
and such large $\epsilon_{\rm B}$ and densities are both at odds with results from afterglow modeling. 

In light of these results, it is difficult to reconcile the maximum synchrotron photon energy predicted by numerical simulations with the observations, which questions our understanding of particle acceleration mechanisms. In shocks in the fast-cooling regime, accelerated electrons lose their energy more rapidly, and hence, even smaller $h\nu_{\rm max}$ would be expected, which intensifies this challenge.
Possible factors that might alleviate or even resolve this tension by allowing for higher maximum energies of electrons accelerated at relativistic shocks are worth mentioning.

The saturation limit of the accelerated electron energy is proportional to the characteristic length scale of turbulence triggered by the Weibel instability, such that $\gamma_{\rm sat} \propto \lambda_{\rm w}$ \citep{ 2014MNRAS.439.2050R, 2022ApJ...925..182H}. This length scale can be expressed as $\lambda_{\rm w}  = \ell_{\rm w} c/\omega_{\rm p}$ \citep{2013ApJ...771...54S, 2014MNRAS.439.2050R}, where $\omega_{\rm p}$ is the relativistic plasma frequency. Hence, the maximum synchrotron photon energy scales as $h \nu_{\rm sat} \propto \ell_{\rm w}^2$. 
Values of $\ell_{\rm w} \sim 10$ were inferred from the numerical simulations 
by \cite{2009ApJ...698.1523S} and \cite{2013ApJ...771...54S} and were adopted in the estimates of 
$h \nu_{\rm sat}$ throughout this work. However, recent  PIC simulations of unprecedented size and duration by
  \citet{2024ApJ...963L..44G}  indicated that the turbulence scale can approach $100 \, c/\omega_{\rm p}$ and may continue to grow beyond the simulated time span. 
The expected $h \nu_{\rm sat}$ can therefore be enhanced by at least two orders of magnitude, and this relaxes the constraints on the parameter space.

Other physical mechanisms might also increase the maximum electron energy beyond
the saturation limit predicted by \citet{2013ApJ...771...54S}.
Using analytical and numerical approaches, \citet{2023MNRAS.522.4955H} and \citet{2023MNRAS.519.1022K} found that 
efficient particle acceleration at relativistic shocks can persist after particles reach the downstream saturation limit as long as their upstream trajectories are dominated by random scattering and not by gyro-motion. 
Furthermore, the three-dimensional structure of preexisting magnetic fields may also come into play 
\citep{2018ApJ...863...18G, 2023MNRAS.522.4955H}. For example, in a toroidal magnetic field configuration, particles with a favorable charge experience a curvature drift along the shock velocity, which can increase the acceleration efficiency and the maximum achievable energy \citep{2023MNRAS.522.4955H}. 

We focused on  XRT afterglow detections at very late times ($> 10^6$ s) under the assumption that the predictions of \citet{2013ApJ...771...54S}, which were gauged for highly relativistic shocks, are still valid. According to their analysis, our conclusions should be even stronger for mildly relativistic shocks with low $\sigma_{\rm u}$, where Weibel instability is still dominant \citep{2026ApJ...998..149J}, which might be the case for the three weakest events in our sample.
Our purpose was to test theoretical predictions based on recent PIC simulations. Therefore, we did not explore alternative theories on particle acceleration, such as the extremely fast acceleration proposed by \citet{2014Sci...343...42A}.

 Clearly, the selection of the six events was influenced by observational constraints and limitations. It is unclear how these biases affect our findings. The requirement of a detectable flux at late times 
might lead to a preferential selection of energetic GRBs or GRBs with particularly late jet break times, however, and thus, as shown, relax the constraints somewhat. The prompt emission energies $E_{\gamma}$ of the six selected GRBs are distributed within a wide range from $\sim 10^{50}$ erg to $> 10^{54}$ erg, which disfavors a selection bias of energetic events.

The jet structure was neglected in our previous discussions. In Fig. ~\ref{fig:parameter} we present the results for two limiting cases of a structured jet. The solid lines correspond to a uniform (top-hat) jet, with a constant energy density within the jet opening angle $\theta_{\rm j}$ and a sharp energy cutoff outside. The dashed lines represent the opposite extreme case, in which $h\nu_{\rm max}$ is always emitted from $\theta_{\rm c}$, related to a spherically symmetric outflow with  a uniform energy distribution.
In reality, the energy distribution is expected to decrease smoothly outside $\theta_{\rm j}$, so the actual constraints should lie between these two cases. 
Previous studies have proposed steep structure jets, where the energy distribution decreases rapidly outside an energetic core (e.g., \citealt{2018MNRAS.478.4128G, 2018PhRvL.120x1103L, 2018ApJ...856L..18M, 2019Sci...363..968G, 2020MNRAS.493.3521B, 2021ApJ...909..114N}). Such models would produce a pronounced jet break \citep{2022MNRAS.515..555B}, and hence, the resulting constraints are expected to be closer to the solid lines in Fig. ~\ref{fig:parameter}. 

To mitigate these limitations and relax the requirement for very late-time observations, further tests of our results and tighter constraints on particle acceleration models in this context, could be achieved by analyzing afterglow spectra at higher photon energies,  above 10 keV, as a spectral cutoff is expected to enter higher-energy bands at earlier times. A less extreme requirement on the detection time might also enlarge the sample of GRBs, which can provide relevant constraints. For instance, NuSTAR can detect photons up to $\sim$ 80 keV, enabling our conclusions to be tested with observations at $\sim 10^{6}$ s. Detections at even earlier epochs, around $\sim 10^{5}$ s, might become feasible with future experiments in the megaelectronvolt band.

\begin{acknowledgements}
This study was funded by the European Union - NextGenerationEU, in the framework of the PRIN Project "PEACE: Powerful Emission and Acceleration in the most powerful Cosmic Explosion" (code 202298J7KT– CUP G53D23000880006). The views and opinions expressed are solely those of the authors and do not necessarily reflect those of the European Union, nor can the European Union be held responsible for them.
We acknowledge financial support from an INAF Mini Grant 2022 (PI L. Nava).
This work made use of data supplied by the UK Swift Science Data Centre at the University of Leicester.
\end{acknowledgements}

\bibliographystyle{aa}
\bibliography{reference}

@ARTICLE{2026ApJ...998..149J,
       author = {{Jikei}, Taiki and {Gro{\v{s}}elj}, Daniel and {Sironi}, Lorenzo},
        title = "{Magnetic Field Amplification and Particle Acceleration in Weakly Magnetized Transrelativistic Electron─Ion Shocks}",
      journal = {\apj},
         year = 2026,
        month = feb,
       volume = {998},
       number = {1},
          eid = {149},
        pages = {149},
          doi = {10.3847/1538-4357/ae3723},
archivePrefix = {arXiv},
       eprint = {2512.03169},
 primaryClass = {astro-ph.HE},
       adsurl = {https://ui.adsabs.harvard.edu/abs/2026ApJ...998..149J}
}

@ARTICLE{2018MNRAS.478.4128G,
       author = {{Gill}, Ramandeep and {Granot}, Jonathan},
        title = "{Afterglow imaging and polarization of misaligned structured GRB jets and cocoons: breaking the degeneracy in GRB 170817A}",
      journal = {\mnras},
         year = 2018,
        month = aug,
       volume = {478},
       number = {3},
        pages = {4128-4141},
          doi = {10.1093/mnras/sty1214},
archivePrefix = {arXiv},
       eprint = {1803.05892},
 primaryClass = {astro-ph.HE},
       adsurl = {https://ui.adsabs.harvard.edu/abs/2018MNRAS.478.4128G}
}

@ARTICLE{2018PhRvL.120x1103L,
       author = {{Lazzati}, Davide and {Perna}, Rosalba and {Morsony}, Brian J. and {Lopez-Camara}, Diego and {Cantiello}, Matteo and {Ciolfi}, Riccardo and {Giacomazzo}, Bruno and {Workman}, Jared C.},
        title = "{Late Time Afterglow Observations Reveal a Collimated Relativistic Jet in the Ejecta of the Binary Neutron Star Merger GW170817}",
      journal = {\prl},
         year = 2018,
        month = jun,
       volume = {120},
       number = {24},
          eid = {241103},
        pages = {241103},
          doi = {10.1103/PhysRevLett.120.241103},
archivePrefix = {arXiv},
       eprint = {1712.03237},
 primaryClass = {astro-ph.HE},
       adsurl = {https://ui.adsabs.harvard.edu/abs/2018PhRvL.120x1103L}
}

@ARTICLE{2019Sci...363..968G,
       author = {{Ghirlanda}, G. and {Salafia}, O.~S. and {Paragi}, Z. and {Giroletti}, M. and {Yang}, J. and {Marcote}, B. and {Blanchard}, J. and {Agudo}, I. and {An}, T. and {Bernardini}, M.~G. and {Beswick}, R. and {Branchesi}, M. and {Campana}, S. and {Casadio}, C. and {Chassande-Mottin}, E. and {Colpi}, M. and {Covino}, S. and {D'Avanzo}, P. and {D'Elia}, V. and {Frey}, S. and {Gawronski}, M. and {Ghisellini}, G. and {Gurvits}, L.~I. and {Jonker}, P.~G. and {van Langevelde}, H.~J. and {Melandri}, A. and {Moldon}, J. and {Nava}, L. and {Perego}, A. and {Perez-Torres}, M.~A. and {Reynolds}, C. and {Salvaterra}, R. and {Tagliaferri}, G. and {Venturi}, T. and {Vergani}, S.~D. and {Zhang}, M.},
        title = "{Compact radio emission indicates a structured jet was produced by a binary neutron star merger}",
      journal = {Science},
         year = 2019,
        month = mar,
       volume = {363},
       number = {6430},
        pages = {968-971},
          doi = {10.1126/science.aau8815},
archivePrefix = {arXiv},
       eprint = {1808.00469},
 primaryClass = {astro-ph.HE},
       adsurl = {https://ui.adsabs.harvard.edu/abs/2019Sci...363..968G}
}

@ARTICLE{2021ApJ...909..114N,
       author = {{Nakar}, Ehud and {Piran}, Tsvi},
        title = "{Afterglow Constraints on the Viewing Angle of Binary Neutron Star Mergers and Determination of the Hubble Constant}",
      journal = {\apj},
         year = 2021,
        month = mar,
       volume = {909},
       number = {2},
          eid = {114},
        pages = {114},
          doi = {10.3847/1538-4357/abd6cd},
archivePrefix = {arXiv},
       eprint = {2005.01754},
 primaryClass = {astro-ph.HE},
       adsurl = {https://ui.adsabs.harvard.edu/abs/2021ApJ...909..114N}
}

@ARTICLE{2018ApJ...856L..18M,
       author = {{Margutti}, R. and {Alexander}, K.~D. and {Xie}, X. and {Sironi}, L. and {Metzger}, B.~D. and {Kathirgamaraju}, A. and {Fong}, W. and {Blanchard}, P.~K. and {Berger}, E. and {MacFadyen}, A. and {Giannios}, D. and {Guidorzi}, C. and {Hajela}, A. and {Chornock}, R. and {Cowperthwaite}, P.~S. and {Eftekhari}, T. and {Nicholl}, M. and {Villar}, V.~A. and {Williams}, P.~K.~G. and {Zrake}, J.},
        title = "{The Binary Neutron Star Event LIGO/Virgo GW170817 160 Days after Merger: Synchrotron Emission across the Electromagnetic Spectrum}",
      journal = {\apjl},
         year = 2018,
        month = mar,
       volume = {856},
       number = {1},
          eid = {L18},
        pages = {L18},
          doi = {10.3847/2041-8213/aab2ad},
archivePrefix = {arXiv},
       eprint = {1801.03531},
 primaryClass = {astro-ph.HE},
       adsurl = {https://ui.adsabs.harvard.edu/abs/2018ApJ...856L..18M}
}

@ARTICLE{2020MNRAS.493.3521B,
       author = {{Beniamini}, Paz and {Granot}, Jonathan and {Gill}, Ramandeep},
        title = "{Afterglow light curves from misaligned structured jets}",
      journal = {\mnras},
         year = 2020,
        month = apr,
       volume = {493},
       number = {3},
        pages = {3521-3534},
          doi = {10.1093/mnras/staa538},
archivePrefix = {arXiv},
       eprint = {2001.02239},
 primaryClass = {astro-ph.HE},
       adsurl = {https://ui.adsabs.harvard.edu/abs/2020MNRAS.493.3521B}
}

@ARTICLE{2022MNRAS.515..555B,
       author = {{Beniamini}, Paz and {Gill}, Ramandeep and {Granot}, Jonathan},
        title = "{Robust features of off-axis gamma-ray burst afterglow light curves}",
      journal = {\mnras},
         year = 2022,
        month = sep,
       volume = {515},
       number = {1},
        pages = {555-570},
          doi = {10.1093/mnras/stac1821},
archivePrefix = {arXiv},
       eprint = {2204.06008},
 primaryClass = {astro-ph.HE},
       adsurl = {https://ui.adsabs.harvard.edu/abs/2022MNRAS.515..555B}
}

@ARTICLE{1979ApJ...228..939C,
       author = {{Cash}, W.},
        title = "{Parameter estimation in astronomy through application of the likelihood ratio.}",
      journal = {\apj},
         year = 1979,
        month = mar,
       volume = {228},
        pages = {939-947},
          doi = {10.1086/156922},
       adsurl = {https://ui.adsabs.harvard.edu/abs/1979ApJ...228..939C}
}

@ARTICLE{BeniaminiNava15,
       author = {{Beniamini}, Paz and {Nava}, Lara and {Duran}, Rodolfo Barniol and {Piran}, Tsvi},
        title = "{Energies of GRB blast waves and prompt efficiencies as implied by modelling of X-ray and GeV afterglows}",
      journal = {\mnras},
         year = 2015,
        month = nov,
       volume = {454},
       number = {1},
        pages = {1073-1085},
          doi = {10.1093/mnras/stv2033},
archivePrefix = {arXiv},
       eprint = {1504.04833},
 primaryClass = {astro-ph.HE},
       adsurl = {https://ui.adsabs.harvard.edu/abs/2015MNRAS.454.1073B}
}

@ARTICLE{1997ApJ...489L..37S,
       author = {{Sari}, Re'em},
        title = "{Hydrodynamics of Gamma-Ray Burst Afterglow}",
      journal = {\apjl},
         year = 1997,
        month = nov,
       volume = {489},
       number = {1},
        pages = {L37-L40},
          doi = {10.1086/310957},
       adsurl = {https://ui.adsabs.harvard.edu/abs/1997ApJ...489L..37S}
}

@ARTICLE{2001ApJ...562L..55F,
       author = {{Frail}, D.~A. and {Kulkarni}, S.~R. and {Sari}, R. and {Djorgovski}, S.~G. and {Bloom}, J.~S. and {Galama}, T.~J. and {Reichart}, D.~E. and {Berger}, E. and {Harrison}, F.~A. and {Price}, P.~A. and {Yost}, S.~A. and {Diercks}, A. and {Goodrich}, R.~W. and {Chaffee}, F.},
        title = "{Beaming in Gamma-Ray Bursts: Evidence for a Standard Energy Reservoir}",
      journal = {\apjl},
         year = 2001,
        month = nov,
       volume = {562},
       number = {1},
        pages = {L55-L58},
          doi = {10.1086/338119},
archivePrefix = {arXiv},
       eprint = {astro-ph/0102282},
 primaryClass = {astro-ph},
       adsurl = {https://ui.adsabs.harvard.edu/abs/2001ApJ...562L..55F}
}

@ARTICLE{2003ApJ...594..674B,
       author = {{Bloom}, J.~S. and {Frail}, D.~A. and {Kulkarni}, S.~R.},
        title = "{Gamma-Ray Burst Energetics and the Gamma-Ray Burst Hubble Diagram: Promises and Limitations}",
      journal = {\apj},
         year = 2003,
        month = sep,
       volume = {594},
       number = {2},
        pages = {674-683},
          doi = {10.1086/377125},
archivePrefix = {arXiv},
       eprint = {astro-ph/0302210},
 primaryClass = {astro-ph},
       adsurl = {https://ui.adsabs.harvard.edu/abs/2003ApJ...594..674B}
}

@ARTICLE{2009ApJ...698...43R,
       author = {{Racusin}, J.~L. and {Liang}, E.~W. and {Burrows}, D.~N. and {Falcone}, A. and {Sakamoto}, T. and {Zhang}, B.~B. and {Zhang}, B. and {Evans}, P. and {Osborne}, J.},
        title = "{Jet Breaks and Energetics of Swift Gamma-Ray Burst X-Ray Afterglows}",
      journal = {\apj},
         year = 2009,
        month = jun,
       volume = {698},
       number = {1},
        pages = {43-74},
          doi = {10.1088/0004-637X/698/1/43},
archivePrefix = {arXiv},
       eprint = {0812.4780},
 primaryClass = {astro-ph},
       adsurl = {https://ui.adsabs.harvard.edu/abs/2009ApJ...698...43R}
}

@ARTICLE{2010ApJ...711..641C,
       author = {{Cenko}, S.~B. and {Frail}, D.~A. and {Harrison}, F.~A. and {Kulkarni}, S.~R. and {Nakar}, E. and {Chandra}, P.~C. and {Butler}, N.~R. and {Fox}, D.~B. and {Gal-Yam}, A. and {Kasliwal}, M.~M. and {Kelemen}, J. and {Moon}, D. -S. and {Ofek}, E.~O. and {Price}, P.~A. and {Rau}, A. and {Soderberg}, A.~M. and {Teplitz}, H.~I. and {Werner}, M.~W. and {Bock}, D.~C. -J. and {Bloom}, J.~S. and {Starr}, D.~A. and {Filippenko}, A.~V. and {Chevalier}, R.~A. and {Gehrels}, N. and {Nousek}, J.~N. and {Piran}, T.},
        title = "{The Collimation and Energetics of the Brightest Swift Gamma-ray Bursts}",
      journal = {\apj},
         year = 2010,
        month = mar,
       volume = {711},
       number = {2},
        pages = {641-654},
          doi = {10.1088/0004-637X/711/2/641},
archivePrefix = {arXiv},
       eprint = {0905.0690},
 primaryClass = {astro-ph.HE},
       adsurl = {https://ui.adsabs.harvard.edu/abs/2010ApJ...711..641C}
}

@ARTICLE{1998ApJ...493L..31P,
       author = {{Panaitescu}, A. and {M{\'e}sz{\'a}ros}, P.},
        title = "{Rings in Fireball Afterglows}",
      journal = {\apjl},
         year = 1998,
        month = jan,
       volume = {493},
       number = {1},
        pages = {L31-L34},
          doi = {10.1086/311127},
archivePrefix = {arXiv},
       eprint = {astro-ph/9709284},
 primaryClass = {astro-ph},
       adsurl = {https://ui.adsabs.harvard.edu/abs/1998ApJ...493L..31P}
}

@ARTICLE{2007ChJAA...7..397H,
       author = {{Huang}, Yong-Feng and {Lu}, Ye and {Wong}, Anna Yuen Lam and {Cheng}, Kwong Sang},
        title = "{A Detailed Study on the Equal Arrival Time Surface Effect in Gamma-Ray Burst Afterglows}",
      journal = {\cjaa},
         year = 2007,
        month = jun,
       volume = {7},
       number = {3},
        pages = {397-404},
          doi = {10.1088/1009-9271/7/3/09},
archivePrefix = {arXiv},
       eprint = {astro-ph/0701846},
 primaryClass = {astro-ph},
       adsurl = {https://ui.adsabs.harvard.edu/abs/2007ChJAA...7..397H}
}

@ARTICLE{2023MNRAS.522.4955H,
       author = {{Huang}, Zhi-Qiu and {Reville}, Brian and {Kirk}, John G. and {Giacinti}, Gwenael},
        title = "{Prospects for ultra-high-energy particle acceleration at relativistic shocks}",
      journal = {\mnras},
         year = 2023,
        month = jul,
       volume = {522},
       number = {4},
        pages = {4955-4962},
          doi = {10.1093/mnras/stad1356},
archivePrefix = {arXiv},
       eprint = {2304.08132},
 primaryClass = {astro-ph.HE},
       adsurl = {https://ui.adsabs.harvard.edu/abs/2023MNRAS.522.4955H}
}

@ARTICLE{2014A&A...564A..77P,
       author = {{Pennanen}, T. and {Vurm}, I. and {Poutanen}, J.},
        title = "{Simulations of gamma-ray burst afterglows with a relativistic kinetic code}",
      journal = {\aap},
         year = 2014,
        month = apr,
       volume = {564},
          eid = {A77},
        pages = {A77},
          doi = {10.1051/0004-6361/201322520},
archivePrefix = {arXiv},
       eprint = {1403.5506},
 primaryClass = {astro-ph.HE},
       adsurl = {https://ui.adsabs.harvard.edu/abs/2014A&A...564A..77P}
}

@ARTICLE{2019ApJ...884..117W,
       author = {{Wang}, Xiang-Yu and {Liu}, Ruo-Yu and {Zhang}, Hai-Ming and {Xi}, Shao-Qiang and {Zhang}, Bing},
        title = "{Synchrotron Self-Compton Emission from External Shocks as the Origin of the Sub-TeV Emission in GRB 180720B and GRB 190114C}",
      journal = {\apj},
         year = 2019,
        month = oct,
       volume = {884},
       number = {2},
          eid = {117},
        pages = {117},
          doi = {10.3847/1538-4357/ab426c},
archivePrefix = {arXiv},
       eprint = {1905.11312},
 primaryClass = {astro-ph.HE},
       adsurl = {https://ui.adsabs.harvard.edu/abs/2019ApJ...884..117W}
}

@ARTICLE{2004ApJ...616..331G,
       author = {{Ghirlanda}, Giancarlo and {Ghisellini}, Gabriele and {Lazzati}, Davide},
        title = "{The Collimation-corrected Gamma-Ray Burst Energies Correlate with the Peak Energy of Their {\ensuremath{\nu}}F$_{{\ensuremath{\nu}}}$ Spectrum}",
      journal = {\apj},
         year = 2004,
        month = nov,
       volume = {616},
       number = {1},
        pages = {331-338},
          doi = {10.1086/424913},
archivePrefix = {arXiv},
       eprint = {astro-ph/0405602},
 primaryClass = {astro-ph},
       adsurl = {https://ui.adsabs.harvard.edu/abs/2004ApJ...616..331G}
}

@ARTICLE{2023MNRAS.518.1522D,
       author = {{Duncan}, Ruby A. and {van der Horst}, Alexander J. and {Beniamini}, Paz},
        title = "{Constraints on electron acceleration in gamma-ray bursts afterglows from radio peaks}",
      journal = {\mnras},
         year = 2023,
        month = jan,
       volume = {518},
       number = {1},
        pages = {1522-1530},
          doi = {10.1093/mnras/stac3172},
archivePrefix = {arXiv},
       eprint = {2211.00686},
 primaryClass = {astro-ph.HE},
       adsurl = {https://ui.adsabs.harvard.edu/abs/2023MNRAS.518.1522D}
}

@ARTICLE{2024MNRAS.527.6752G,
       author = {{Garcia-Cifuentes}, Keneth and {Becerra}, Rosa Leticia and {De Colle}, Fabio and {Vargas}, Felipe},
        title = "{Unraveling parameter degeneracy in GRB data analysis}",
      journal = {\mnras},
         year = 2024,
        month = jan,
       volume = {527},
       number = {3},
        pages = {6752-6762},
          doi = {10.1093/mnras/stad3625},
archivePrefix = {arXiv},
       eprint = {2309.15825},
 primaryClass = {astro-ph.HE},
       adsurl = {https://ui.adsabs.harvard.edu/abs/2024MNRAS.527.6752G}
}

@ARTICLE{2023ApJ...956...12I,
       author = {{Isravel}, Hebzibha and {B{\'e}gu{\'e}}, Damien and {Pe'er}, Asaf},
        title = "{Hybrid Emission Modeling of GRB 221009A: Shedding Light on TeV Emission Origins in Long GRBs}",
      journal = {\apj},
         year = 2023,
        month = oct,
       volume = {956},
       number = {1},
          eid = {12},
        pages = {12},
          doi = {10.3847/1538-4357/acefcd},
archivePrefix = {arXiv},
       eprint = {2308.06994},
 primaryClass = {astro-ph.HE},
       adsurl = {https://ui.adsabs.harvard.edu/abs/2023ApJ...956...12I}
}

@ARTICLE{2023ApJ...946L..27A,
       author = {{Aharonian}, F. and {Ait Benkhali}, F. and {Aschersleben}, J. and {Ashkar}, H. and {Backes}, M. and {Baktash}, A. and {Barbosa Martins}, V. and {Batzofin}, R. and {Becherini}, Y. and {Berge}, D. and {Bernl{\"o}hr}, K. and {Bi}, B. and {B{\"o}ttcher}, M. and {Boisson}, C. and {Bolmont}, J. and {de Bony de Lavergne}, M. and {Borowska}, J. and {Bouyahiaoui}, M. and {Bradascio}, F. and {Breuhaus}, M. and {Brose}, R. and {Brun}, F. and {Bruno}, B. and {Bulik}, T. and {Burger-Scheidlin}, C. and {Caroff}, S. and {Casanova}, S. and {Celic}, J. and {Cerruti}, M. and {Chand}, T. and {Chandra}, S. and {Chen}, A. and {Chibueze}, J. and {Chibueze}, O. and {Cotter}, G. and {Dai}, S. and {Damascene Mbarubucyeye}, J. and {Devin}, J. and {Djannati-Ata{\"\i}}, A. and {Dmytriiev}, A. and {Doroshenko}, V. and {Egberts}, K. and {Einecke}, S. and {Ernenwein}, J.-P. and {Fegan}, S. and {Fichet de Clairfontaine}, G. and {Filipovic}, M. and {Fontaine}, G. and {F{\"u}{\ss}ling}, M. and {Funk}, S. and {Gabici}, S. and {Ghafourizadeh}, S. and {Giavitto}, G. and {Glawion}, D. and {Glicenstein}, J.~F. and {Goswami}, P. and {Grolleron}, G. and {Grondin}, M.-H. and {Hinton}, J.~A. and {Holch}, T.~L. and {Holler}, M. and {Horns}, D. and {Huang}, Zhiqiu and {Jamrozy}, M. and {Jankowsky}, F. and {Joshi}, V. and {Jung-Richardt}, I. and {Kasai}, E. and {Katarzy{\'n}ski}, K. and {Khatoon}, R. and {Kh{\'e}lifi}, B. and {Klu{\'z}niak}, W. and {Komin}, Nu. and {Konno}, R. and {Kosack}, K. and {Kostunin}, D. and {Lang}, R.~G. and {Le Stum}, S. and {Leitl}, F. and {Lemi{\`e}re}, A. and {Lemoine-Goumard}, M. and {Lenain}, J.-P. and {Leuschner}, F. and {Lohse}, T. and {Lypova}, I. and {Mackey}, J. and {Malyshev}, D. and {Malyshev}, D. and {Marandon}, V. and {Marchegiani}, P. and {Marcowith}, A. and {Mart{\'\i}-Devesa}, G. and {Marx}, R. and {Meyer}, M. and {Mitchell}, A. and {Mohrmann}, L. and {Montanari}, A. and {Moulin}, E. and {Murach}, T. and {Nakashima}, K. and {de Naurois}, M. and {Niemiec}, J. and {Noel}, A. Priyana and {O'Brien}, P. and {Ohm}, S. and {Olivera-Nieto}, L. and {de Ona Wilhelmi}, E. and {Ostrowski}, M. and {Panny}, S. and {Panter}, M. and {Parsons}, R.~D. and {Peron}, G. and {Prokhorov}, D.~A. and {Prokoph}, H. and {P{\"u}hlhofer}, G. and {Punch}, M. and {Quirrenbach}, A. and {Reichherzer}, P. and {Reimer}, A. and {Reimer}, O. and {Ren}, H. and {Renaud}, M. and {Reville}, B. and {Rieger}, F. and {Rowell}, G. and {Rudak}, B. and {Ruiz-Velasco}, E. and {Sahakian}, V. and {Salzmann}, H. and {Santangelo}, A. and {Sasaki}, M. and {Sch{\"a}fer}, J. and {Sch{\"u}ssler}, F. and {Schutte}, H.~M. and {Schwanke}, U. and {Shapopi}, J.~N.~S. and {Specovius}, A. and {Spencer}, S. and {Stawarz}, {\L}. and {Steenkamp}, R. and {Steinmassl}, S. and {Steppa}, C. and {Sushch}, I. and {Suzuki}, H. and {Takahashi}, T. and {Tanaka}, T. and {Terrier}, R. and {Tsuji}, N. and {Uchiyama}, Y. and {Vecchi}, M. and {Venter}, C. and {Vink}, J. and {Wagner}, S.~J. and {White}, R. and {Wierzcholska}, A. and {Wong}, Yu Wun and {Zacharias}, M. and {Zargaryan}, D. and {Zdziarski}, A.~A. and {Zech}, A. and {Zhu}, S.~J. and {{\.Z}ywucka}, N. and {H.~E.~S.~S. Collaboration}},
        title = "{H.E.S.S. Follow-up Observations of GRB 221009A}",
      journal = {\apjl},
         year = 2023,
        month = mar,
       volume = {946},
       number = {1},
          eid = {L27},
        pages = {L27},
          doi = {10.3847/2041-8213/acc405},
archivePrefix = {arXiv},
       eprint = {2303.10558},
 primaryClass = {astro-ph.HE},
       adsurl = {https://ui.adsabs.harvard.edu/abs/2023ApJ...946L..27A}
}

@ARTICLE{2023MNRAS.519.1022K,
       author = {{Kirk}, John G. and {Reville}, Brian and {Huang}, Zhi-Qiu},
        title = "{Particle acceleration at ultrarelativistic, perpendicular shock fronts}",
      journal = {\mnras},
         year = 2023,
        month = feb,
       volume = {519},
       number = {1},
        pages = {1022-1029},
          doi = {10.1093/mnras/stac3589},
archivePrefix = {arXiv},
       eprint = {2212.02349},
 primaryClass = {physics.plasm-ph},
       adsurl = {https://ui.adsabs.harvard.edu/abs/2023MNRAS.519.1022K}
}

@ARTICLE{2018ApJ...863...18G,
       author = {{Giacinti}, Gwenael and {Kirk}, John G.},
        title = "{Acceleration of X-Ray Emitting Electrons in the Crab Nebula}",
      journal = {\apj},
         year = 2018,
        month = aug,
       volume = {863},
       number = {1},
          eid = {18},
        pages = {18},
          doi = {10.3847/1538-4357/aacffb},
archivePrefix = {arXiv},
       eprint = {1804.05056},
 primaryClass = {astro-ph.HE},
       adsurl = {https://ui.adsabs.harvard.edu/abs/2018ApJ...863...18G}
}

@ARTICLE{2009MNRAS.397.1177E,
       author = {{Evans}, P.~A. and {Beardmore}, A.~P. and {Page}, K.~L. and {Osborne}, J.~P. and {O'Brien}, P.~T. and {Willingale}, R. and {Starling}, R.~L.~C. and {Burrows}, D.~N. and {Godet}, O. and {Vetere}, L. and {Racusin}, J. and {Goad}, M.~R. and {Wiersema}, K. and {Angelini}, L. and {Capalbi}, M. and {Chincarini}, G. and {Gehrels}, N. and {Kennea}, J.~A. and {Margutti}, R. and {Morris}, D.~C. and {Mountford}, C.~J. and {Pagani}, C. and {Perri}, M. and {Romano}, P. and {Tanvir}, N.},
        title = "{Methods and results of an automatic analysis of a complete sample of Swift-XRT observations of GRBs}",
      journal = {\mnras},
         year = 2009,
        month = aug,
       volume = {397},
       number = {3},
        pages = {1177-1201},
          doi = {10.1111/j.1365-2966.2009.14913.x},
archivePrefix = {arXiv},
       eprint = {0812.3662},
 primaryClass = {astro-ph},
       adsurl = {https://ui.adsabs.harvard.edu/abs/2009MNRAS.397.1177E}
}

@ARTICLE{2001ApJ...548..787S,
       author = {{Sari}, Re'em and {Esin}, Ann A.},
        title = "{On the Synchrotron Self-Compton Emission from Relativistic Shocks and Its Implications for Gamma-Ray Burst Afterglows}",
      journal = {\apj},
         year = 2001,
        month = feb,
       volume = {548},
       number = {2},
        pages = {787-799},
          doi = {10.1086/319003},
archivePrefix = {arXiv},
       eprint = {astro-ph/0005253},
 primaryClass = {astro-ph},
       adsurl = {https://ui.adsabs.harvard.edu/abs/2001ApJ...548..787S}
}

@ARTICLE{1976PhFl...19.1130B,
       author = {{Blandford}, R.~D. and {McKee}, C.~F.},
        title = "{Fluid dynamics of relativistic blast waves}",
      journal = {Physics of Fluids},
         year = 1976,
        month = aug,
       volume = {19},
        pages = {1130-1138},
          doi = {10.1063/1.861619},
       adsurl = {https://ui.adsabs.harvard.edu/abs/1976PhFl...19.1130B}
}

@ARTICLE{2022GCN.32648....1D,
       author = {{de Ugarte Postigo}, A. and {Izzo}, L. and {Pugliese}, G. and {Xu}, D. and {Schneider}, B. and {Fynbo}, J.~P.~U. and {Tanvir}, N.~R. and {Malesani}, D.~B. and {Saccardi}, A. and {Kann}, D.~A. and {Wiersema}, K. and {Gompertz}, B.~P. and {Thoene}, C.~C. and {Levan}, A.~J. and {Stargate Collaboration}},
        title = "{GRB 221009A: Redshift from X-shooter/VLT}",
      journal = {GRB Coordinates Network},
         year = 2022,
        month = oct,
       volume = {32648},
        pages = {1},
       adsurl = {https://ui.adsabs.harvard.edu/abs/2022GCN.32648....1D}
}

@ARTICLE{2022GCN.32668....1F,
       author = {{Frederiks}, D. and {Lysenko}, A. and {Ridnaia}, A. and {Svinkin}, D. and {Tsvetkova}, A. and {Ulanov}, M. and {Cline}, T. and {Konus-Wind Team}},
        title = "{Konus-Wind detection of GRB 221009A}",
      journal = {GRB Coordinates Network},
         year = 2022,
        month = oct,
       volume = {32668},
        pages = {1},
       adsurl = {https://ui.adsabs.harvard.edu/abs/2022GCN.32668....1F}
}

@ARTICLE{2022ApJ...925..182H,
       author = {{Huang}, Zhi-Qiu and {Kirk}, John G. and {Giacinti}, Gwenael and {Reville}, Brian},
        title = "{The Implications of TeV-detected GRB Afterglows for Acceleration at Relativistic Shocks}",
      journal = {\apj},
         year = 2022,
        month = feb,
       volume = {925},
       number = {2},
          eid = {182},
        pages = {182},
          doi = {10.3847/1538-4357/ac3f38},
archivePrefix = {arXiv},
       eprint = {2112.00111},
 primaryClass = {astro-ph.HE},
       adsurl = {https://ui.adsabs.harvard.edu/abs/2022ApJ...925..182H}
}

@ARTICLE{2019GCN.25660....1T,
       author = {{Tsvetkova}, A. and {Golenetskii}, S. and {Aptekar}, R. and {Frederiks}, D. and {Ulanov}, M. and {Svinkin}, D. and {Lysenko}, A. and {Kozlova}, A. and {Cline}, T. and {Konus-Wind Team}},
        title = "{Konus-Wind observation of GRB 190829A}",
      journal = {GRB Coordinates Network},
         year = 2019,
        month = sep,
       volume = {25660},
        pages = {1},
       adsurl = {https://ui.adsabs.harvard.edu/abs/2019GCN.25660....1T}
}

@ARTICLE{2019GCN.25575....1L,
       author = {{Lesage}, S. and {Poolakkil}, S. and {Fletcher}, C. and {Meegan}, C. and {Goldstein}, A. and {Fermi GBM Team}},
        title = "{GRB 190829A: Fermi GBM detection}",
      journal = {GRB Coordinates Network},
         year = 2019,
        month = aug,
       volume = {25575},
        pages = {1},
       adsurl = {https://ui.adsabs.harvard.edu/abs/2019GCN.25575....1L}
}

@ARTICLE{2019GCN.25565....1V,
       author = {{Valeev}, A.~F. and {Castro-Tirado}, A.~J. and {Hu}, Y. -D. and {Fernandez-Garcia}, E. and {Sokolov}, V.~V. and {Carrasco}, I. and {Castellon}, A. and {Garcia Alvarez}, D. and {Rivero}, M. and {et al.}},
        title = "{GRB 190829A: 10.4m GTC spectroscopy}",
      journal = {GRB Coordinates Network},
         year = 2019,
        month = aug,
       volume = {25565},
        pages = {1},
       adsurl = {https://ui.adsabs.harvard.edu/abs/2019GCN.25565....1V}
}

@ARTICLE{2016GCN.20323....1F,
       author = {{Frederiks}, D. and {Golenetskii}, S. and {Aptekar}, R. and {Oleynik}, P. and {Ulanov}, M. and {Svinkin}, D. and {Tsvetkova}, A. and {Lysenko}, A. and {Kozlova}, A. and {Cline}, T.},
        title = "{Konus-Wind observation of GRB 161219B.}",
      journal = {GRB Coordinates Network},
         year = 2016,
        month = jan,
       volume = {20323},
        pages = {1},
       adsurl = {https://ui.adsabs.harvard.edu/abs/2016GCN.20323....1F}
}

@ARTICLE{2016GCN.20321....1T,
       author = {{Tanvir}, N.~R. and {Kruehler}, T. and {Wiersema}, K. and {Xu}, D. and {Malesani}, D. and {Milvang-Jensen}, B. and {Fynbo}, J.~P.~U.},
        title = "{GRB 161219B: VLT/X-shooter redshift.}",
      journal = {GRB Coordinates Network},
         year = 2016,
        month = jan,
       volume = {20321},
        pages = {1},
       adsurl = {https://ui.adsabs.harvard.edu/abs/2016GCN.20321....1T}
}

@ARTICLE{2013ApJ...771...54S,
       author = {{Sironi}, Lorenzo and {Spitkovsky}, Anatoly and {Arons}, Jonathan},
        title = "{The Maximum Energy of Accelerated Particles in Relativistic Collisionless Shocks}",
      journal = {\apj},
         year = 2013,
        month = jul,
       volume = {771},
       number = {1},
          eid = {54},
        pages = {54},
          doi = {10.1088/0004-637X/771/1/54},
archivePrefix = {arXiv},
       eprint = {1301.5333},
 primaryClass = {astro-ph.HE},
       adsurl = {https://ui.adsabs.harvard.edu/abs/2013ApJ...771...54S}
}

@ARTICLE{2013GCN.15260....1G,
       author = {{Golenetskii}, S. and {Aptekar}, R. and {Frederiks}, D. and {Pal'Shin}, V. and {Oleynik}, P. and {Ulanov}, M. and {Svinkin}, D. and {Cline}, T.},
        title = "{Konus-wind observation of GRB 130925A.}",
      journal = {GRB Coordinates Network},
         year = 2013,
        month = jan,
       volume = {15260},
        pages = {1},
       adsurl = {https://ui.adsabs.harvard.edu/abs/2013GCN.15260....1G}
}

@ARTICLE{2013GCN.14986....1G,
       author = {{Golenetskii}, S. and {Aptekar}, R. and {Pal'Shin}, V. and {Frederiks}, D. and {Oleynik}, P. and {Ulanov}, M. and {Svinkin}, D. and {Cline}, T.},
        title = "{Konus-wind observation of GRB 130702A.}",
      journal = {GRB Coordinates Network},
         year = 2013,
        month = jan,
       volume = {14986},
        pages = {1},
       adsurl = {https://ui.adsabs.harvard.edu/abs/2013GCN.14986....1G}
}

@ARTICLE{2013GCN.14455....1L,
       author = {{Levan}, A.~J. and {Cenko}, S.~B. and {Perley}, D.~A. and {Tanvir}, N.~R.},
        title = "{GRB 130427A: gemini-north redshift.}",
      journal = {GRB Coordinates Network},
         year = 2013,
        month = jan,
       volume = {14455},
        pages = {1},
       adsurl = {https://ui.adsabs.harvard.edu/abs/2013GCN.14455....1L}
}

@ARTICLE{2013GCN.15249....1V,
       author = {{Vreeswijk}, P.~M. and {Malesani}, D. and {Fynbo}, J.~P.~U. and {De Cia}, A. and {Ledoux}, C.},
        title = "{GRB 130925A: VLT/UVES observations.}",
      journal = {GRB Coordinates Network},
         year = 2013,
        month = jan,
       volume = {15249},
        pages = {1},
       adsurl = {https://ui.adsabs.harvard.edu/abs/2013GCN.15249....1V}
}

@ARTICLE{2013GCN.14983....1L,
       author = {{Leloudas}, G. and {Fynbo}, J.~P.~U. and {Schulze}, S. and {Xu}, D. and {Malesani}, D. and {Geier}, S. and {Cano}, Z. and {Jakobsson}, P.},
        title = "{GRB 130702A: NOT spectroscopy and redshift of the nearby bright galaxy.}",
      journal = {GRB Coordinates Network},
         year = 2013,
        month = jan,
       volume = {14983},
        pages = {1},
       adsurl = {https://ui.adsabs.harvard.edu/abs/2013GCN.14983....1L}
}

@ARTICLE{2002ApJ...568..820G,
       author = {{Granot}, Jonathan and {Sari}, Re'em},
        title = "{The Shape of Spectral Breaks in Gamma-Ray Burst Afterglows}",
      journal = {\apj},
         year = 2002,
        month = apr,
       volume = {568},
       number = {2},
        pages = {820-829},
          doi = {10.1086/338966},
archivePrefix = {arXiv},
       eprint = {astro-ph/0108027},
 primaryClass = {astro-ph},
       adsurl = {https://ui.adsabs.harvard.edu/abs/2002ApJ...568..820G}
}

@ARTICLE{1998ApJ...494L..49S,
       author = {{Sari}, Re'em},
        title = "{The Observed Size and Shape of Gamma-Ray Burst Afterglow}",
      journal = {\apjl},
         year = 1998,
        month = feb,
       volume = {494},
       number = {1},
        pages = {L49-L52},
          doi = {10.1086/311160},
archivePrefix = {arXiv},
       eprint = {astro-ph/9709300},
 primaryClass = {astro-ph},
       adsurl = {https://ui.adsabs.harvard.edu/abs/1998ApJ...494L..49S}
}

@ARTICLE{2007A&A...469..379E,
       author = {{Evans}, P.~A. and {Beardmore}, A.~P. and {Page}, K.~L. and {Tyler}, L.~G. and {Osborne}, J.~P. and {Goad}, M.~R. and {O'Brien}, P.~T. and {Vetere}, L. and {Racusin}, J. and {Morris}, D. and {Burrows}, D.~N. and {Capalbi}, M. and {Perri}, M. and {Gehrels}, N. and {Romano}, P.},
        title = "{An online repository of Swift/XRT light curves of {\ensuremath{\gamma}}-ray bursts}",
      journal = {\aap},
         year = 2007,
        month = jul,
       volume = {469},
       number = {1},
        pages = {379-385},
          doi = {10.1051/0004-6361:20077530},
archivePrefix = {arXiv},
       eprint = {0704.0128},
 primaryClass = {astro-ph},
       adsurl = {https://ui.adsabs.harvard.edu/abs/2007A&A...469..379E}
}

@ARTICLE{1996ApJ...457..253D,
       author = {{de Jager}, O.~C. and {Harding}, A.~K. and {Michelson}, P.~F. and {Nel}, H.~I. and {Nolan}, P.~L. and {Sreekumar}, P. and {Thompson}, D.~J.},
        title = "{Gamma-Ray Observations of the Crab Nebula: A Study of the Synchro-Compton Spectrum}",
      journal = {\apj},
         year = 1996,
        month = jan,
       volume = {457},
        pages = {253},
          doi = {10.1086/176726},
       adsurl = {https://ui.adsabs.harvard.edu/abs/1996ApJ...457..253D}
}

@BOOK{2009herb.book.....D,
       author = {{Dermer}, Charles D. and {Menon}, Govind},
        title = "{High Energy Radiation from Black Holes: Gamma Rays, Cosmic Rays, and Neutrinos}",
        publisher = {Princeton University Press},
         year = 2009,
       adsurl = {https://ui.adsabs.harvard.edu/abs/2009herb.book.....D}
}

@ARTICLE{2007ApJ...655..989Z,
       author = {{Zhang}, Bing and {Liang}, Enwei and {Page}, Kim L. and {Grupe}, Dirk and {Zhang}, Bin-Bin and {Barthelmy}, Scott D. and {Burrows}, David N. and {Campana}, Sergio and {Chincarini}, Guido and {Gehrels}, Neil and {Kobayashi}, Shiho and {M{\'e}sz{\'a}ros}, Peter and {Moretti}, Alberto and {Nousek}, John A. and {O'Brien}, Paul T. and {Osborne}, Julian P. and {Roming}, Peter W.~A. and {Sakamoto}, Takanori and {Schady}, Patricia and {Willingale}, Richard},
        title = "{GRB Radiative Efficiencies Derived from the Swift Data: GRBs versus XRFs, Long versus Short}",
      journal = {\apj},
         year = 2007,
        month = feb,
       volume = {655},
       number = {2},
        pages = {989-1001},
          doi = {10.1086/510110},
archivePrefix = {arXiv},
       eprint = {astro-ph/0610177},
 primaryClass = {astro-ph},
       adsurl = {https://ui.adsabs.harvard.edu/abs/2007ApJ...655..989Z}
}

@ARTICLE{2015ApJS..219....9W,
       author = {{Wang}, Xiang-Gao and {Zhang}, Bing and {Liang}, En-Wei and {Gao}, He and {Li}, Liang and {Deng}, Can-Min and {Qin}, Song-Mei and {Tang}, Qing-Wen and {Kann}, D. Alexander and {Ryde}, Felix and {Kumar}, Pawan},
        title = "{How Bad or Good Are the External Forward Shock Afterglow Models of Gamma-Ray Bursts?}",
      journal = {\apjs},
         year = 2015,
        month = jul,
       volume = {219},
       number = {1},
          eid = {9},
        pages = {9},
          doi = {10.1088/0067-0049/219/1/9},
archivePrefix = {arXiv},
       eprint = {1503.03193},
 primaryClass = {astro-ph.HE},
       adsurl = {https://ui.adsabs.harvard.edu/abs/2015ApJS..219....9W}
}

@ARTICLE{2004ApJ...613..477L,
       author = {{Lloyd-Ronning}, Nicole M. and {Zhang}, Bing},
        title = "{On the Kinetic Energy and Radiative Efficiency of Gamma-Ray Bursts}",
      journal = {\apj},
         year = 2004,
        month = sep,
       volume = {613},
       number = {1},
        pages = {477-483},
          doi = {10.1086/423026},
archivePrefix = {arXiv},
       eprint = {astro-ph/0404107},
 primaryClass = {astro-ph},
       adsurl = {https://ui.adsabs.harvard.edu/abs/2004ApJ...613..477L}
}

@ARTICLE{2012MNRAS.425..506D,
       author = {{D'Avanzo}, P. and {Salvaterra}, R. and {Sbarufatti}, B. and {Nava}, L. and {Melandri}, A. and {Bernardini}, M.~G. and {Campana}, S. and {Covino}, S. and {Fugazza}, D. and {Ghirlanda}, G. and {Ghisellini}, G. and {La Parola}, V. and {Perri}, M. and {Vergani}, S.~D. and {Tagliaferri}, G.},
        title = "{A complete sample of bright Swift Gamma-ray bursts: X-ray afterglow luminosity and its correlation with the prompt emission}",
      journal = {\mnras},
         year = 2012,
        month = sep,
       volume = {425},
       number = {1},
        pages = {506-513},
          doi = {10.1111/j.1365-2966.2012.21489.x},
archivePrefix = {arXiv},
       eprint = {1206.2357},
 primaryClass = {astro-ph.HE},
       adsurl = {https://ui.adsabs.harvard.edu/abs/2012MNRAS.425..506D}
}

@ARTICLE{2024ApJ...972..195L,
       author = {{Li}, Liang and {Wang}, Yu},
        title = "{GRB 190114C: Fireball Energy Budget and Radiative Efficiency Revisited}",
      journal = {\apj},
         year = 2024,
        month = sep,
       volume = {972},
       number = {2},
          eid = {195},
        pages = {195},
          doi = {10.3847/1538-4357/ad2511},
archivePrefix = {arXiv},
       eprint = {2302.06116},
 primaryClass = {astro-ph.HE},
       adsurl = {https://ui.adsabs.harvard.edu/abs/2024ApJ...972..195L}
}

@ARTICLE{2009ApJ...698.1523S,
       author = {{Sironi}, Lorenzo and {Spitkovsky}, Anatoly},
        title = "{Particle Acceleration in Relativistic Magnetized Collisionless Pair Shocks: Dependence of Shock Acceleration on Magnetic Obliquity}",
      journal = {\apj},
         year = 2009,
        month = jun,
       volume = {698},
       number = {2},
        pages = {1523-1549},
          doi = {10.1088/0004-637X/698/2/1523},
archivePrefix = {arXiv},
       eprint = {0901.2578},
 primaryClass = {astro-ph.HE},
       adsurl = {https://ui.adsabs.harvard.edu/abs/2009ApJ...698.1523S}
}

@ARTICLE{2014MNRAS.439.2050R,
       author = {{Reville}, B. and {Bell}, A.~R.},
        title = "{On the maximum energy of shock-accelerated cosmic rays at ultra-relativistic shocks}",
      journal = {\mnras},
         year = 2014,
        month = apr,
       volume = {439},
       number = {2},
        pages = {2050-2059},
          doi = {10.1093/mnras/stu088},
archivePrefix = {arXiv},
       eprint = {1401.2803},
 primaryClass = {astro-ph.HE},
       adsurl = {https://ui.adsabs.harvard.edu/abs/2014MNRAS.439.2050R}
}

@ARTICLE{2024ApJ...963L..44G,
       author = {{Gro{\v{s}}elj}, Daniel and {Sironi}, Lorenzo and {Spitkovsky}, Anatoly},
        title = "{Long-term Evolution of Relativistic Unmagnetized Collisionless Shocks}",
      journal = {\apjl},
         year = 2024,
        month = mar,
       volume = {963},
       number = {2},
          eid = {L44},
        pages = {L44},
          doi = {10.3847/2041-8213/ad2c8c},
archivePrefix = {arXiv},
       eprint = {2401.02392},
 primaryClass = {astro-ph.HE},
       adsurl = {https://ui.adsabs.harvard.edu/abs/2024ApJ...963L..44G}
}

@ARTICLE{2012ApJ...749...80S,
       author = {{Sagi}, Eran and {Nakar}, Ehud},
        title = "{On Particle Acceleration Rate in Gamma-Ray Burst Afterglows}",
      journal = {\apj},
         year = 2012,
        month = apr,
       volume = {749},
       number = {1},
          eid = {80},
        pages = {80},
          doi = {10.1088/0004-637X/749/1/80},
archivePrefix = {arXiv},
       eprint = {1201.5124},
 primaryClass = {astro-ph.HE},
       adsurl = {https://ui.adsabs.harvard.edu/abs/2012ApJ...749...80S}
}

@ARTICLE{2013GCN.14487....1G,
       author = {{Golenetskii}, S. and {Aptekar}, R. and {Frederiks}, D. and {Mazets}, E. and {Pal'Shin}, V. and {Oleynik}, P. and {Ulanov}, M. and {Svinkin}, D. and {Cline}, T.},
        title = "{Konus-wind observation of GRB 130427A.}",
      journal = {GRB Coordinates Network},
         year = 2013,
        month = jan,
       volume = {14487},
        pages = {1},
       adsurl = {https://ui.adsabs.harvard.edu/abs/2013GCN.14487....1G}
}

@ARTICLE{2010ApJ...718L..63P,
       author = {{Piran}, Tsvi and {Nakar}, Ehud},
        title = "{On the External Shock Synchrotron Model for Gamma-ray Bursts' GeV Emission}",
      journal = {\apjl},
         year = 2010,
        month = aug,
       volume = {718},
       number = {2},
        pages = {L63-L67},
          doi = {10.1088/2041-8205/718/2/L63},
archivePrefix = {arXiv},
       eprint = {1003.5919},
 primaryClass = {astro-ph.HE},
       adsurl = {https://ui.adsabs.harvard.edu/abs/2010ApJ...718L..63P}
}

@ARTICLE{2022ApJ...931L..19S,
       author = {{Salafia}, Om Sharan and {Ravasio}, Maria Edvige and {Yang}, Jun and {An}, Tao and {Orienti}, Monica and {Ghirlanda}, Giancarlo and {Nava}, Lara and {Giroletti}, Marcello and {Mohan}, Prashanth and {Spinelli}, Riccardo and {Zhang}, Yingkang and {Marcote}, Benito and {Cim{\`o}}, Giuseppe and {Wu}, Xuefeng and {Li}, Zhixuan},
        title = "{Multiwavelength View of the Close-by GRB 190829A Sheds Light on Gamma-Ray Burst Physics}",
      journal = {\apjl},
         year = 2022,
        month = jun,
       volume = {931},
       number = {2},
          eid = {L19},
        pages = {L19},
          doi = {10.3847/2041-8213/ac6c28},
archivePrefix = {arXiv},
       eprint = {2106.07169},
 primaryClass = {astro-ph.HE},
       adsurl = {https://ui.adsabs.harvard.edu/abs/2022ApJ...931L..19S}
}

@ARTICLE{2014Sci...343...42A,
       author = {{Ackermann}, M. and {Ajello}, M. and {Asano}, K. and {Atwood}, W.~B. and {Axelsson}, M. and {Baldini}, L. and {Ballet}, J. and {Barbiellini}, G. and {Baring}, M.~G. and {Bastieri}, D. and {Bechtol}, K. and {Bellazzini}, R. and {Bissaldi}, E. and {Bonamente}, E. and {Bregeon}, J. and {Brigida}, M. and {Bruel}, P. and {Buehler}, R. and {Burgess}, J. Michael and {Buson}, S. and {Caliandro}, G.~A. and {Cameron}, R.~A. and {Caraveo}, P.~A. and {Cecchi}, C. and {Chaplin}, V. and {Charles}, E. and {Chekhtman}, A. and {Cheung}, C.~C. and {Chiang}, J. and {Chiaro}, G. and {Ciprini}, S. and {Claus}, R. and {Cleveland}, W. and {Cohen-Tanugi}, J. and {Collazzi}, A. and {Cominsky}, L.~R. and {Connaughton}, V. and {Conrad}, J. and {Cutini}, S. and {D'Ammando}, F. and {de Angelis}, A. and {DeKlotz}, M. and {de Palma}, F. and {Dermer}, C.~D. and {Desiante}, R. and {Diekmann}, A. and {Di Venere}, L. and {Drell}, P.~S. and {Drlica-Wagner}, A. and {Favuzzi}, C. and {Fegan}, S.~J. and {Ferrara}, E.~C. and {Finke}, J. and {Fitzpatrick}, G. and {Focke}, W.~B. and {Franckowiak}, A. and {Fukazawa}, Y. and {Funk}, S. and {Fusco}, P. and {Gargano}, F. and {Gehrels}, N. and {Germani}, S. and {Gibby}, M. and {Giglietto}, N. and {Giles}, M. and {Giordano}, F. and {Giroletti}, M. and {Godfrey}, G. and {Granot}, J. and {Grenier}, I.~A. and {Grove}, J.~E. and {Gruber}, D. and {Guiriec}, S. and {Hadasch}, D. and {Hanabata}, Y. and {Harding}, A.~K. and {Hayashida}, M. and {Hays}, E. and {Horan}, D. and {Hughes}, R.~E. and {Inoue}, Y. and {Jogler}, T. and {J{\'o}hannesson}, G. and {Johnson}, W.~N. and {Kawano}, T. and {Kn{\"o}dlseder}, J. and {Kocevski}, D. and {Kuss}, M. and {Lande}, J. and {Larsson}, S. and {Latronico}, L. and {Longo}, F. and {Loparco}, F. and {Lovellette}, M.~N. and {Lubrano}, P. and {Mayer}, M. and {Mazziotta}, M.~N. and {McEnery}, J.~E. and {Michelson}, P.~F. and {Mizuno}, T. and {Moiseev}, A.~A. and {Monzani}, M.~E. and {Moretti}, E. and {Morselli}, A. and {Moskalenko}, I.~V. and {Murgia}, S. and {Nemmen}, R. and {Nuss}, E. and {Ohno}, M. and {Ohsugi}, T. and {Okumura}, A. and {Omodei}, N. and {Orienti}, M. and {Paneque}, D. and {Pelassa}, V. and {Perkins}, J.~S. and {Pesce-Rollins}, M. and {Petrosian}, V. and {Piron}, F. and {Pivato}, G. and {Porter}, T.~A. and {Racusin}, J.~L. and {Rain{\`o}}, S. and {Rando}, R. and {Razzano}, M. and {Razzaque}, S. and {Reimer}, A. and {Reimer}, O. and {Ritz}, S. and {Roth}, M. and {Ryde}, F. and {Sartori}, A. and {Parkinson}, P.~M. Saz and {Scargle}, J.~D. and {Schulz}, A. and {Sgr{\`o}}, C. and {Siskind}, E.~J. and {Sonbas}, E. and {Spandre}, G. and {Spinelli}, P. and {Tajima}, H. and {Takahashi}, H. and {Thayer}, J.~G. and {Thayer}, J.~B. and {Thompson}, D.~J. and {Tibaldo}, L. and {Tinivella}, M. and {Torres}, D.~F. and {Tosti}, G. and {Troja}, E. and {Usher}, T.~L. and {Vandenbroucke}, J. and {Vasileiou}, V. and {Vianello}, G. and {Vitale}, V. and {Winer}, B.~L. and {Wood}, K.~S. and {Yamazaki}, R. and {Younes}, G. and {Yu}, H. -F. and {Zhu}, S.~J. and {Bhat}, P.~N. and {Briggs}, M.~S. and {Byrne}, D. and {Foley}, S. and {Goldstein}, A. and {Jenke}, P. and {Kippen}, R.~M. and {Kouveliotou}, C. and {McBreen}, S. and {Meegan}, C. and {Paciesas}, W.~S. and {Preece}, R. and {Rau}, A. and {Tierney}, D. and {van der Horst}, A.~J. and {von Kienlin}, A. and {Wilson-Hodge}, C. and {Xiong}, S. and {Cusumano}, G. and {La Parola}, V. and {Cummings}, J.~R.},
        title = "{Fermi-LAT Observations of the Gamma-Ray Burst GRB 130427A}",
      journal = {Science},
         year = 2014,
        month = jan,
       volume = {343},
       number = {6166},
        pages = {42-47},
          doi = {10.1126/science.1242353},
archivePrefix = {arXiv},
       eprint = {1311.5623},
 primaryClass = {astro-ph.HE},
       adsurl = {https://ui.adsabs.harvard.edu/abs/2014Sci...343...42A}
}

@ARTICLE{2016MNRAS.462.2990D,
       author = {{Du}, Shuang and {L{\"u}}, Hou-Jun and {Zhong}, Shu-Qing and {Liang}, En-Wei},
        title = "{The radiative efficiency of relativistic jet and wind: a case study of GRB 070110}",
      journal = {\mnras},
         year = 2016,
        month = nov,
       volume = {462},
       number = {3},
        pages = {2990-2994},
          doi = {10.1093/mnras/stw1869},
archivePrefix = {arXiv},
       eprint = {1607.08324},
 primaryClass = {astro-ph.HE},
       adsurl = {https://ui.adsabs.harvard.edu/abs/2016MNRAS.462.2990D}
}

@ARTICLE{2014MNRAS.442...20H,
       author = {{Hasco{\"e}t}, R. and {Daigne}, F. and {Mochkovitch}, R.},
        title = "{The prompt-early afterglow connection in gamma-ray bursts: implications for the early afterglow physics}",
      journal = {\mnras},
         year = 2014,
        month = jul,
       volume = {442},
       number = {1},
        pages = {20-27},
          doi = {10.1093/mnras/stu750},
archivePrefix = {arXiv},
       eprint = {1401.0751},
 primaryClass = {astro-ph.HE},
       adsurl = {https://ui.adsabs.harvard.edu/abs/2014MNRAS.442...20H}
}

@ARTICLE{2018ApJS..236...26L,
       author = {{Li}, Liang and {Wu}, Xue-Feng and {Lei}, Wei-Hua and {Dai}, Zi-Gao and {Liang}, En-Wei and {Ryde}, Felix},
        title = "{Constraining the Type of Central Engine of GRBs with Swift Data}",
      journal = {\apjs},
         year = 2018,
        month = jun,
       volume = {236},
       number = {2},
          eid = {26},
        pages = {26},
          doi = {10.3847/1538-4365/aabaf3},
archivePrefix = {arXiv},
       eprint = {1712.09390},
 primaryClass = {astro-ph.HE},
       adsurl = {https://ui.adsabs.harvard.edu/abs/2018ApJS..236...26L}
}

@ARTICLE{1990ARA&A..28..215D,
       author = {{Dickey}, John M. and {Lockman}, Felix J.},
        title = "{H I in the galaxy.}",
      journal = {\araa},
         year = 1990,
        month = jan,
       volume = {28},
        pages = {215-261},
          doi = {10.1146/annurev.aa.28.090190.001243},
       adsurl = {https://ui.adsabs.harvard.edu/abs/1990ARA&A..28..215D}
}

@ARTICLE{2005A&A...440..775K,
       author = {{Kalberla}, P.~M.~W. and {Burton}, W.~B. and {Hartmann}, Dap and {Arnal}, E.~M. and {Bajaja}, E. and {Morras}, R. and {P{\"o}ppel}, W.~G.~L.},
        title = "{The Leiden/Argentine/Bonn (LAB) Survey of Galactic HI. Final data release of the combined LDS and IAR surveys with improved stray-radiation corrections}",
      journal = {\aap},
         year = 2005,
        month = sep,
       volume = {440},
       number = {2},
        pages = {775-782},
          doi = {10.1051/0004-6361:20041864},
archivePrefix = {arXiv},
       eprint = {astro-ph/0504140},
 primaryClass = {astro-ph},
       adsurl = {https://ui.adsabs.harvard.edu/abs/2005A&A...440..775K}
}

@ARTICLE{2016A&A...594A.116H,
       author = {{HI4PI Collaboration} and {Ben Bekhti}, N. and {Fl{\"o}er}, L. and {Keller}, R. and {Kerp}, J. and {Lenz}, D. and {Winkel}, B. and {Bailin}, J. and {Calabretta}, M.~R. and {Dedes}, L. and {Ford}, H.~A. and {Gibson}, B.~K. and {Haud}, U. and {Janowiecki}, S. and {Kalberla}, P.~M.~W. and {Lockman}, F.~J. and {McClure-Griffiths}, N.~M. and {Murphy}, T. and {Nakanishi}, H. and {Pisano}, D.~J. and {Staveley-Smith}, L.},
        title = "{HI4PI: A full-sky H I survey based on EBHIS and GASS}",
      journal = {\aap},
         year = 2016,
        month = oct,
       volume = {594},
          eid = {A116},
        pages = {A116},
          doi = {10.1051/0004-6361/201629178},
archivePrefix = {arXiv},
       eprint = {1610.06175},
 primaryClass = {astro-ph.GA},
       adsurl = {https://ui.adsabs.harvard.edu/abs/2016A&A...594A.116H}
}

@ARTICLE{2019Natur.575..459M,
       author = {{MAGIC Collaboration} and {Acciari}, V.~A. and {Ansoldi}, S. and {Antonelli}, L.~A. and {Engels}, A. Arbet and {Baack}, D. and {Babi{\'c}}, A. and {Banerjee}, B. and {Barres de Almeida}, U. and {Barrio}, J.~A. and {Becerra Gonz{\'a}lez}, J. and {Bednarek}, W. and {Bellizzi}, L. and {Bernardini}, E. and {Berti}, A. and {Besenrieder}, J. and {Bhattacharyya}, W. and {Bigongiari}, C. and {Biland}, A. and {Blanch}, O. and {Bonnoli}, G. and {Bo{\v{s}}njak}, {\v{Z}}. and {Busetto}, G. and {Carosi}, R. and {Ceribella}, G. and {Chai}, Y. and {Chilingaryan}, A. and {Cikota}, S. and {Colak}, S.~M. and {Colin}, U. and {Colombo}, E. and {Contreras}, J.~L. and {Cortina}, J. and {Covino}, S. and {D'Elia}, V. and {da Vela}, P. and {Dazzi}, F. and {de Angelis}, A. and {de Lotto}, B. and {Delfino}, M. and {Delgado}, J. and {Depaoli}, D. and {di Pierro}, F. and {di Venere}, L. and {Do Souto Espi{\~n}eira}, E. and {Dominis Prester}, D. and {Donini}, A. and {Dorner}, D. and {Doro}, M. and {Elsaesser}, D. and {Fallah Ramazani}, V. and {Fattorini}, A. and {Ferrara}, G. and {Fidalgo}, D. and {Foffano}, L. and {Fonseca}, M.~V. and {Font}, L. and {Fruck}, C. and {Fukami}, S. and {Garc{\'\i}a L{\'o}pez}, R.~J. and {Garczarczyk}, M. and {Gasparyan}, S. and {Gaug}, M. and {Giglietto}, N. and {Giordano}, F. and {Godinovi{\'c}}, N. and {Green}, D. and {Guberman}, D. and {Hadasch}, D. and {Hahn}, A. and {Herrera}, J. and {Hoang}, J. and {Hrupec}, D. and {H{\"u}tten}, M. and {Inada}, T. and {Inoue}, S. and {Ishio}, K. and {Iwamura}, Y. and {Jouvin}, L. and {Kerszberg}, D. and {Kubo}, H. and {Kushida}, J. and {Lamastra}, A. and {Lelas}, D. and {Leone}, F. and {Lindfors}, E. and {Lombardi}, S. and {Longo}, F. and {L{\'o}pez}, M. and {L{\'o}pez-Coto}, R. and {L{\'o}pez-Oramas}, A. and {Loporchio}, S. and {Machado de Oliveira Fraga}, B. and {Maggio}, C. and {Majumdar}, P. and {Makariev}, M. and {Mallamaci}, M. and {Maneva}, G. and {Manganaro}, M. and {Mannheim}, K. and {Maraschi}, L. and {Mariotti}, M. and {Mart{\'\i}nez}, M. and {Mazin}, D. and {Mi{\'c}anovi{\'c}}, S. and {Miceli}, D. and {Minev}, M. and {Miranda}, J.~M. and {Mirzoyan}, R. and {Molina}, E. and {Moralejo}, A. and {Morcuende}, D. and {Moreno}, V. and {Moretti}, E. and {Munar-Adrover}, P. and {Neustroev}, V. and {Nigro}, C. and {Nilsson}, K. and {Ninci}, D. and {Nishijima}, K. and {Noda}, K. and {Nogu{\'e}s}, L. and {Nozaki}, S. and {Paiano}, S. and {Palatiello}, M. and {Paneque}, D. and {Paoletti}, R. and {Paredes}, J.~M. and {Pe{\~n}il}, P. and {Peresano}, M. and {Persic}, M. and {Moroni}, P.~G. Prada and {Prandini}, E. and {Puljak}, I. and {Rhode}, W. and {Rib{\'o}}, M. and {Rico}, J. and {Righi}, C. and {Rugliancich}, A. and {Saha}, L. and {Sahakyan}, N. and {Saito}, T. and {Sakurai}, S. and {Satalecka}, K. and {Schmidt}, K. and {Schweizer}, T. and {Sitarek}, J. and {{\v{S}}nidari{\'c}}, I. and {Sobczynska}, D. and {Somero}, A. and {Stamerra}, A. and {Strom}, D. and {Strzys}, M. and {Suda}, Y. and {Suri{\'c}}, T. and {Takahashi}, M. and {Tavecchio}, F. and {Temnikov}, P. and {Terzi{\'c}}, T. and {Teshima}, M. and {Torres-Alb{\`a}}, N. and {Tosti}, L. and {Vagelli}, V. and {van Scherpenberg}, J. and {Vanzo}, G. and {Vazquez Acosta}, M. and {Vigorito}, C.~F. and {Vitale}, V. and {Vovk}, I. and {Will}, M. and {Zari{\'c}}, D. and {Nava}, L. and {Veres}, P. and {Bhat}, P.~N. and {Briggs}, M.~S. and {Cleveland}, W.~H. and {Hamburg}, R. and {Hui}, C.~M. and {Mailyan}, B. and {Preece}, R.~D. and {Roberts}, O.~J. and {von Kienlin}, A. and {Wilson-Hodge}, C.~A. and {Kocevski}, D. and {Arimoto}, M. and {Tak}, D. and {Asano}, K. and {Axelsson}, M. and {Barbiellini}, G. and {Bissaldi}, E. and {Dirirsa}, F. Fana and {Gill}, R. and {Granot}, J. and {McEnery}, J. and {Omodei}, N. and {Razzaque}, S. and {Piron}, F. and {Racusin}, J.~L. and {Thompson}, D.~J. and {Campana}, S.},
        title = "{Observation of inverse Compton emission from a long {\ensuremath{\gamma}}-ray burst}",
      journal = {\nat},
         year = 2019,
        month = nov,
       volume = {575},
       number = {7783},
        pages = {459-463},
          doi = {10.1038/s41586-019-1754-6},
archivePrefix = {arXiv},
       eprint = {2006.07251},
 primaryClass = {astro-ph.HE},
       adsurl = {https://ui.adsabs.harvard.edu/abs/2019Natur.575..459M}
}

@ARTICLE{2023MNRAS.518..174E,
       author = {{Evans}, P.~A. and {Page}, K.~L. and {Beardmore}, A.~P. and {Eyles-Ferris}, R.~A.~J. and {Osborne}, J.~P. and {Campana}, S. and {Kennea}, J.~A. and {Cenko}, S.~B.},
        title = "{A real-time transient detector and the living Swift-XRT point source catalogue}",
      journal = {\mnras},
         year = 2023,
        month = jan,
       volume = {518},
       number = {1},
        pages = {174-184},
          doi = {10.1093/mnras/stac2937},
archivePrefix = {arXiv},
       eprint = {2208.14478},
 primaryClass = {astro-ph.HE},
       adsurl = {https://ui.adsabs.harvard.edu/abs/2023MNRAS.518..174E}
}

@ARTICLE{1999ApJ...526..697M,
       author = {{Medvedev}, Mikhail V. and {Loeb}, Abraham},
        title = "{Generation of Magnetic Fields in the Relativistic Shock of Gamma-Ray Burst Sources}",
      journal = {\apj},
         year = 1999,
        month = dec,
       volume = {526},
       number = {2},
        pages = {697-706},
          doi = {10.1086/308038},
archivePrefix = {arXiv},
       eprint = {astro-ph/9904363},
 primaryClass = {astro-ph},
       adsurl = {https://ui.adsabs.harvard.edu/abs/1999ApJ...526..697M}
}

@ARTICLE{1959PhRvL...2...83W,
       author = {{Weibel}, Erich S.},
        title = "{Spontaneously Growing Transverse Waves in a Plasma Due to an Anisotropic Velocity Distribution}",
      journal = {\prl},
         year = 1959,
        month = feb,
       volume = {2},
       number = {3},
        pages = {83-84},
          doi = {10.1103/PhysRevLett.2.83},
       adsurl = {https://ui.adsabs.harvard.edu/abs/1959PhRvL...2...83W}
}

@INPROCEEDINGS{2005AIPC..801..345S,
       author = {{Spitkovsky}, Anatoly},
        title = "{Simulations of relativistic collisionless shocks: shock structure and particle acceleration}",
    booktitle = {Astrophysical Sources of High Energy Particles and Radiation},
         year = 2005,
       editor = {{Bulik}, Tomasz and {Rudak}, Bronislaw and {Madejski}, Grzegorz},
       series = {American Institute of Physics Conference Series},
       volume = {801},
        month = nov,
    publisher = {AIP},
        pages = {345-350},
          doi = {10.1063/1.2141897},
archivePrefix = {arXiv},
       eprint = {astro-ph/0603211},
 primaryClass = {astro-ph},
       adsurl = {https://ui.adsabs.harvard.edu/abs/2005AIPC..801..345S}
}

@ARTICLE{2009ApJ...707L..92S,
       author = {{Sironi}, Lorenzo and {Spitkovsky}, Anatoly},
        title = "{Synthetic Spectra from Particle-In-Cell Simulations of Relativistic Collisionless Shocks}",
      journal = {\apjl},
         year = 2009,
        month = dec,
       volume = {707},
       number = {1},
        pages = {L92-L96},
          doi = {10.1088/0004-637X/707/1/L92},
archivePrefix = {arXiv},
       eprint = {0908.3193},
 primaryClass = {astro-ph.HE},
       adsurl = {https://ui.adsabs.harvard.edu/abs/2009ApJ...707L..92S}
}

@ARTICLE{2008ApJ...673L..39S,
       author = {{Spitkovsky}, Anatoly},
        title = "{On the Structure of Relativistic Collisionless Shocks in Electron-Ion Plasmas}",
      journal = {\apjl},
         year = 2008,
        month = jan,
       volume = {673},
       number = {1},
        pages = {L39},
          doi = {10.1086/527374},
archivePrefix = {arXiv},
       eprint = {0706.3126},
 primaryClass = {astro-ph},
       adsurl = {https://ui.adsabs.harvard.edu/abs/2008ApJ...673L..39S}
}

@ARTICLE{1998NewA....3..157D,
       author = {{Dermer}, Charles D. and {Chiang}, James},
        title = "{Electron acceleration and synchrotron radiation in decelerating plasmoids}",
      journal = {\na},
         year = 1998,
        month = apr,
       volume = {3},
       number = {3},
        pages = {157-173},
          doi = {10.1016/S1384-1076(98)00004-9},
archivePrefix = {arXiv},
       eprint = {astro-ph/9712052},
 primaryClass = {astro-ph},
       adsurl = {https://ui.adsabs.harvard.edu/abs/1998NewA....3..157D}
}

@ARTICLE{1994ApJ...432..181M,
       author = {{Meszaros}, P. and {Rees}, M.~J. and {Papathanassiou}, H.},
        title = "{Spectral Properties of Blast-Wave Models of Gamma-Ray Burst Sources}",
      journal = {\apj},
         year = 1994,
        month = sep,
       volume = {432},
        pages = {181},
          doi = {10.1086/174559},
archivePrefix = {arXiv},
       eprint = {astro-ph/9311071},
 primaryClass = {astro-ph},
       adsurl = {https://ui.adsabs.harvard.edu/abs/1994ApJ...432..181M}
}

@ARTICLE{1998ApJ...497L..17S,
       author = {{Sari}, Re'em and {Piran}, Tsvi and {Narayan}, Ramesh},
        title = "{Spectra and Light Curves of Gamma-Ray Burst Afterglows}",
      journal = {\apjl},
         year = 1998,
        month = apr,
       volume = {497},
       number = {1},
        pages = {L17-L20},
          doi = {10.1086/311269},
archivePrefix = {arXiv},
       eprint = {astro-ph/9712005},
 primaryClass = {astro-ph},
       adsurl = {https://ui.adsabs.harvard.edu/abs/1998ApJ...497L..17S}
}

@ARTICLE{1999ApJ...512..699C,
       author = {{Chiang}, James and {Dermer}, Charles D.},
        title = "{Synchrotron and Synchrotron Self-Compton Emission and the Blast-Wave Model of Gamma-Ray Bursts}",
      journal = {\apj},
         year = 1999,
        month = feb,
       volume = {512},
       number = {2},
        pages = {699-710},
          doi = {10.1086/306789},
archivePrefix = {arXiv},
       eprint = {astro-ph/9803339},
 primaryClass = {astro-ph},
       adsurl = {https://ui.adsabs.harvard.edu/abs/1999ApJ...512..699C}
}

@ARTICLE{2012ApJ...756..189F,
       author = {{Fong}, W. and {Berger}, E. and {Margutti}, R. and {Zauderer}, B.~A. and {Troja}, E. and {Czekala}, I. and {Chornock}, R. and {Gehrels}, N. and {Sakamoto}, T. and {Fox}, D.~B. and {Podsiadlowski}, P.},
        title = "{A Jet Break in the X-Ray Light Curve of Short GRB 111020A: Implications for Energetics and Rates}",
      journal = {\apj},
         year = 2012,
        month = sep,
       volume = {756},
       number = {2},
          eid = {189},
        pages = {189},
          doi = {10.1088/0004-637X/756/2/189},
archivePrefix = {arXiv},
       eprint = {1204.5475},
 primaryClass = {astro-ph.HE},
       adsurl = {https://ui.adsabs.harvard.edu/abs/2012ApJ...756..189F}
}

@ARTICLE{Chevalier2000,
       author = {{Chevalier}, Roger A. and {Li}, Zhi-Yun},
        title = "{Wind Interaction Models for Gamma-Ray Burst Afterglows: The Case for Two Types of Progenitors}",
      journal = {\apj},
         year = 2000,
        month = jun,
       volume = {536},
       number = {1},
        pages = {195-212},
          doi = {10.1086/308914},
archivePrefix = {arXiv},
       eprint = {astro-ph/9908272},
 primaryClass = {astro-ph},
       adsurl = {https://ui.adsabs.harvard.edu/abs/2000ApJ...536..195C}
}

@ARTICLE{Gao2013,
       author = {{Gao}, He and {Lei}, Wei-Hua and {Zou}, Yuan-Chuan and {Wu}, Xue-Feng and {Zhang}, Bing},
        title = "{A complete reference of the analytical synchrotron external shock models of gamma-ray bursts}",
      journal = {\nar},
         year = 2013,
        month = dec,
       volume = {57},
       number = {6},
        pages = {141-190},
          doi = {10.1016/j.newar.2013.10.001},
archivePrefix = {arXiv},
       eprint = {1310.2181},
 primaryClass = {astro-ph.HE},
       adsurl = {https://ui.adsabs.harvard.edu/abs/2013NewAR..57..141G}
}

@ARTICLE{Rhoads1999,
       author = {{Rhoads}, James E.},
        title = "{The Dynamics and Light Curves of Beamed Gamma-Ray Burst Afterglows}",
      journal = {\apj},
         year = 1999,
        month = nov,
       volume = {525},
       number = {2},
        pages = {737-749},
          doi = {10.1086/307907},
archivePrefix = {arXiv},
       eprint = {astro-ph/9903399},
 primaryClass = {astro-ph},
       adsurl = {https://ui.adsabs.harvard.edu/abs/1999ApJ...525..737R}
}

@ARTICLE{Wu2026,
       author = {{Wu}, Zhao-Feng and {Guevara-Montoya}, Sof{\'\i}a and {Beniamini}, Paz and {Giannios}, Dimitrios and {Gro{\v{s}}elj}, Daniel and {Sironi}, Lorenzo},
        title = "{Maximum Energy of Particles Accelerated in GRB Afterglow Shocks}",
      journal = {arXiv e-prints},
         year = 2026,
        month = jan,
          eid = {arXiv:2601.19135},
        pages = {arXiv:2601.19135},
          doi = {10.48550/arXiv.2601.19135},
archivePrefix = {arXiv},
       eprint = {2601.19135},
 primaryClass = {astro-ph.HE},
       adsurl = {https://ui.adsabs.harvard.edu/abs/2026arXiv260119135W}
}

@ARTICLE{Banerjee2025,
       author = {{Banerjee}, Biswajit and {Macera}, Samanta and {De Santis}, Alessio L. and {Mei}, Alessio and {Tissino}, Jacopo and {Oganesyan}, Gor and {Frederiks}, Dmitry D. and {Lysenko}, Alexandra L. and {Svinkin}, Dmitry S. and {Tsvetkova}, Anastasia E. and {Branchesi}, Marica},
        title = "{Observation of the spectral turnover in the afterglow emission of GRB 221009A}",
      journal = {\aap},
         year = 2025,
        month = sep,
       volume = {701},
          eid = {A68},
        pages = {A68},
          doi = {10.1051/0004-6361/202554813},
archivePrefix = {arXiv},
       eprint = {2405.15855},
 primaryClass = {astro-ph.HE},
       adsurl = {https://ui.adsabs.harvard.edu/abs/2025A&A...701A..68B}
}

\begin{appendix}

\section{Fitting results}
\label{appendix:fit}

\begin{figure*}
\centering
\includegraphics[scale=0.4]{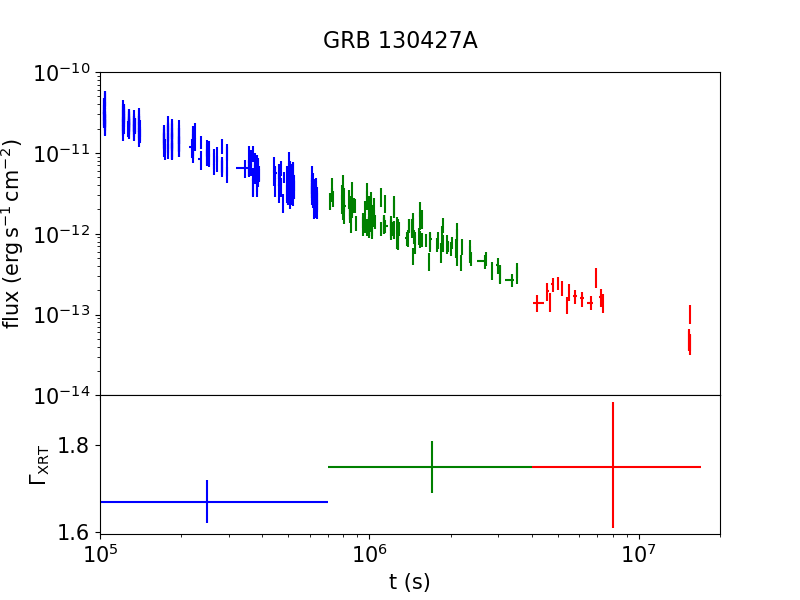}
\includegraphics[scale=0.4]{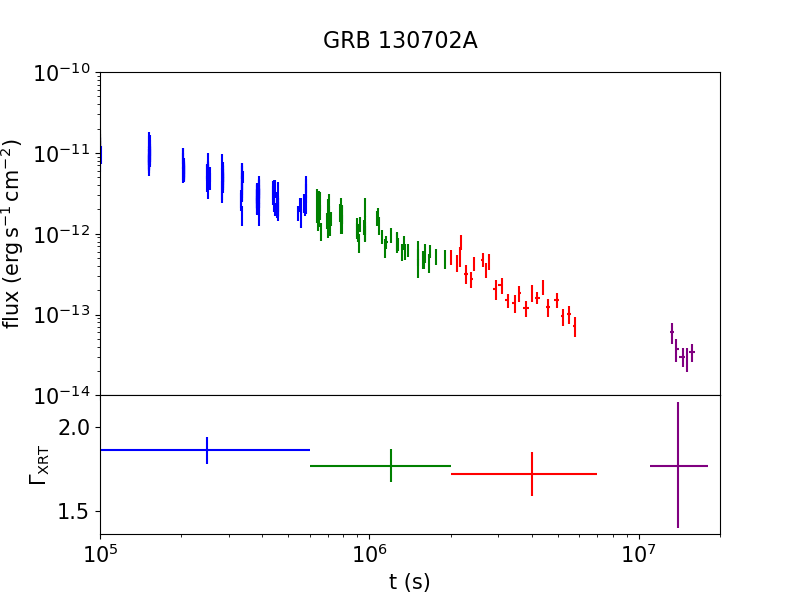}
\includegraphics[scale=0.4]{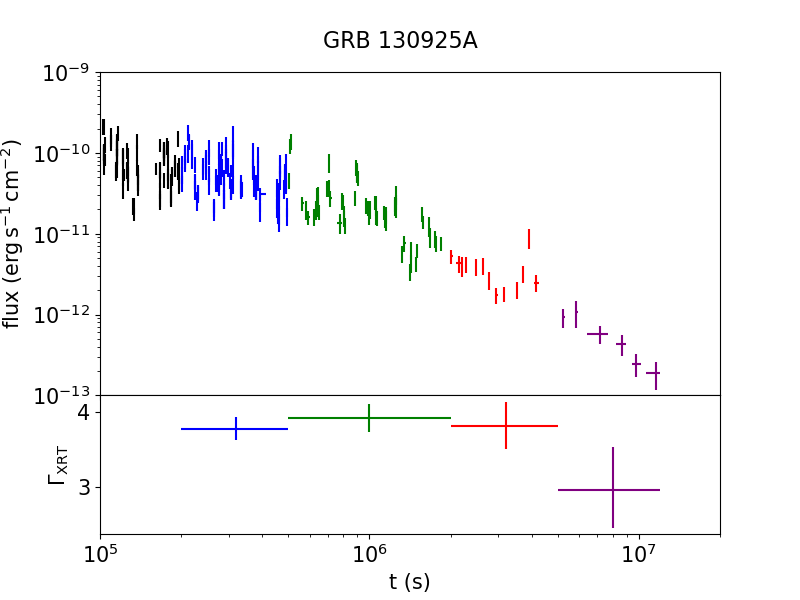}
\includegraphics[scale=0.4]{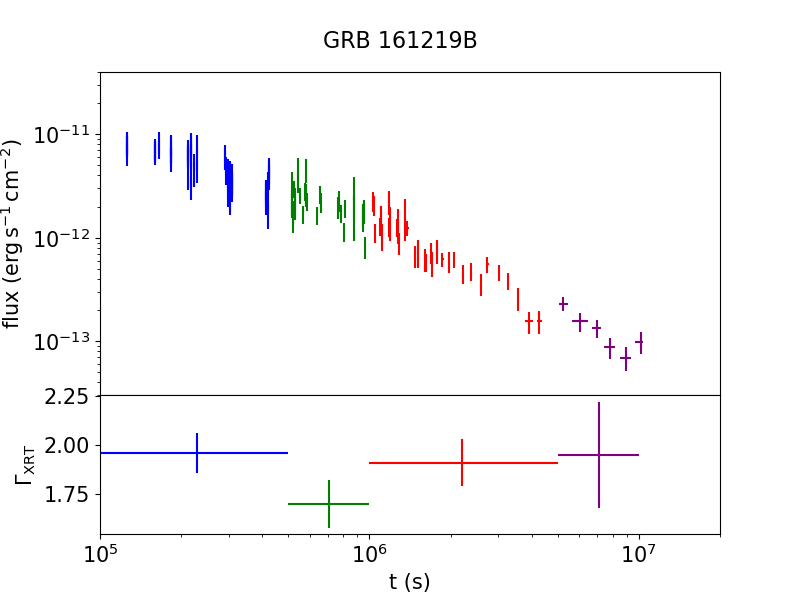}
\includegraphics[scale=0.4]{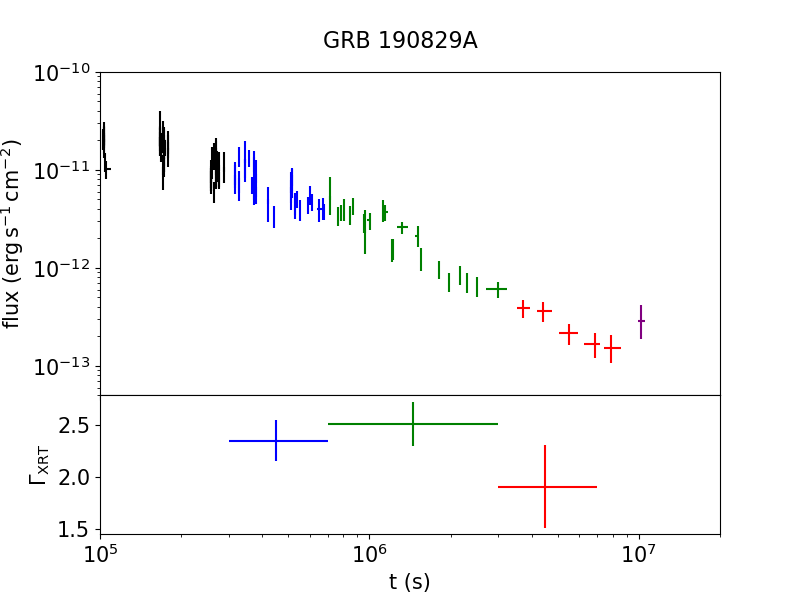}
\includegraphics[scale=0.4]{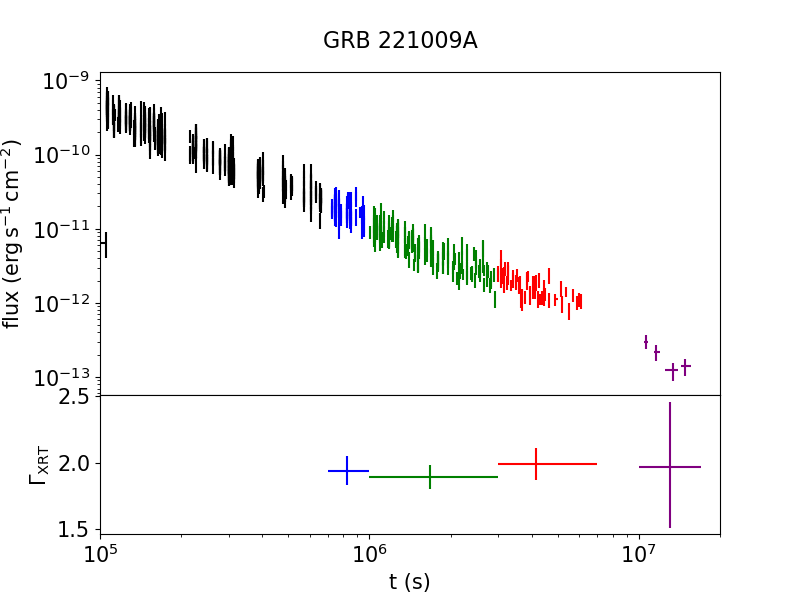}
\caption{The top panels show the late-time ($\gtrsim 10^{5}$ s) unabsorbed X-ray (0.3–10 keV) light curves of the six GRBs analyzed in this work, obtained from the
\emph{Swift}/XRT repository. 
 Different colors indicate the time intervals selected for the spectral analysis. The bottom panels display the corresponding photon indices derived from single power-law spectral fits and listed in Tab.~\ref{tab:mcmc}). 
 \label{fig:index}}
\end{figure*}

Figure~\ref{fig:index} shows the unabsorbed XRT light curves in the 0.3–10 keV energy band for the six GRBs at times $\gtrsim 10^{5}$ s. Different colors indicate the time intervals selected for the spectral analysis. The lower panels display the corresponding time evolution of the photon indices $\Gamma_{\rm XRT}$ derived from single power-law fits. If $h\nu_{\rm max}$ enters the X-ray band, a spectral cutoff is expected, leading to a softening of the X-ray spectra. However, our analysis shows that, for each GRB, the photon indices measured at different time intervals are mutually consistent within the uncertainties. 

The fitting parameters obtained from the MCMC analysis assuming a cutoff power-law model
are listed in Tab.~\ref{tab:mcmc}. The parameter $\log (E_{\rm cut}/ {\rm keV})$ in the last column of Tab.~\ref{tab:mcmc} are fitted within the range [-1, 2]. In each time interval, the upper bound of $\log (E_{\rm cut}/ {\rm keV})$ is not well convergent. Therefore we adopt the lower bound of the one-sided 95$\%$ credible interval as the lower limit of the cutoff energy.

\begin{table*}[ht]
\centering
\caption[]{Parameters used to fit the six selected GRBs with the cutoff power-law model.
\label{tab:mcmc}}
\renewcommand{\arraystretch}{1.8}
\begin{tabular}{|c|c|c|c|c|c|c|c|}
\hline
GRB                      & $z$                     & t$^{a}$  (s)    & $N_{\rm H, Gal}^b$            & $N_{\rm H}^c$          & $\log_{10}$Flux$^d$ (cgs)   & $\Gamma_{\rm XRT}^{e}$                 & $\log_{10}$($E_{\rm cut}$/keV)$^f$         \\ 
                  &                     &  & ($\times \, 10^{22} \, \rm{cm^{-2}})$             & ($\times \, 10^{22}\, \rm{cm^{-2}})$             & &             &        \\ \hline
                  
\multirow{4}{*}{221009A} & \multirow{4}{*}{0.151}  & $7 \times 10^5$ - $10^6$ & \multirow{4}{*}{0.443}  & \multirow{4}{*}{0.604} & $-10.88^{+0.02}_{-0.02}$ & \multirow{4}{*}{$1.84^{+0.07}_{-0.11}$} & \textgreater 1.16 \\ \cline{3-3} \cline{6-6} \cline{8-8} 
                         &                         & $10^6$ - $3 \times 10^6$ &                         &                        &$-11.40^{+0.02}_{-0.02}$&                       & \textgreater 1.30 \\ \cline{3-3} \cline{6-6} \cline{8-8} 
                         &                         & $3\times 10^6$ - $7\times 10^6$ &                         &                        & $-11.90^{+0.03}_{-0.03}$ &                       & \textgreater 1.03 \\ \cline{3-3} \cline{6-6} \cline{8-8} 
                         &                         & $10^7$ - $2 \times 10^7$ &                         &                        & $-13.02^{+0.10}_{-0.11}$ &                       & \textgreater 0.77 \\ \hline
\multirow{3}{*}{190829A} & \multirow{3}{*}{0.0785} & $3\times 10^5$ - $7\times 10^5$ & \multirow{3}{*}{0.0466} & \multirow{3}{*}{0.749} & $-11.36^{+0.04}_{-0.05}$ & \multirow{3}{*}{$2.29^{+0.15}_{-0.21}$} & \textgreater 1.04 \\ \cline{3-3} \cline{6-6} \cline{8-8} 
                         &                         & $7\times 10^5$ - $3\times 10^6$ &                         &                        & $-11.94^{+0.05}_{-0.05}$ &                       & \textgreater 0.82 \\ \cline{3-3} \cline{6-6} \cline{8-8} 
                         &                         & $3\times 10^6$ - $10^7$ &                         &                        & $-12.76^{+0.10}_{-0.10}$ &                       & \textgreater 1.11 \\ \hline
\multirow{4}{*}{161219B} & \multirow{4}{*}{0.1475} & $10^5$ - $5\times 10^5$ & \multirow{4}{*}{0.0294} & \multirow{4}{*}{0.172} & $-11.33^{+0.03}_{-0.03}$ & \multirow{4}{*}{$1.76^{+0.10}_{-0.11}$} & \textgreater 0.86 \\ \cline{3-3} \cline{6-6} \cline{8-8} 
                         &                         & $5\times 10^5$ - $10^6$ &                         &                        & $-11.71^{+0.04}_{-0.03}$ &                       & \textgreater 1.41 \\ \cline{3-3} \cline{6-6} \cline{8-8} 
                         &                         & $10^6$ - $5\times 10^6$ &                         &                        & $-12.24^{+0.04}_{-0.04}$ &                       & \textgreater 0.94 \\ \cline{3-3} \cline{6-6} \cline{8-8} 
                         &                         & $5\times 10^6$ - $2\times 10^7$ &                         &                        & $-12.93^{+0.08}_{-0.09}$ &                       & \textgreater 0.71 \\ \hline
\multirow{4}{*}{130925A} & \multirow{4}{*}{0.347}  & $2\times 10^5$ - $5\times 10^5$ & \multirow{4}{*}{0.016}  & \multirow{4}{*}{1.962} & $-10.41^{+0.06}_{-0.06}$ & \multirow{4}{*}{$3.72^{+0.11}_{-0.18}$} & \textgreater 0.95 \\ \cline{3-3} \cline{6-6} \cline{8-8} 
                         &                         & $5\times 10^5$ - $2\times 10^6$ &                         &                        & $-11.08^{+0.06}_{-0.06}$ &                       & \textgreater 0.83 \\ \cline{3-3} \cline{6-6} \cline{8-8} 
                         &                         & $2\times 10^6$ - $5\times 10^6$ &                         &                        & $-11.72^{+0.09}_{-0.10}$ &                       & \textgreater 0.81 \\ \cline{3-3} \cline{6-6} \cline{8-8} 
                         &                         & $5\times 10^6$ - $2\times 10^7$ &                         &                        & $-12.43^{+0.13}_{-0.14}$ &                       & \textgreater 0.94 \\ \hline
\multirow{4}{*}{130702A} & \multirow{4}{*}{0.145}  & $1\times 10^5$ - $6\times 10^5$ & \multirow{4}{*}{0.0149} & \multirow{4}{*}{0.085} & $-11.38^{+0.03}_{-0.03}$ & \multirow{4}{*}{$1.73^{+0.08}_{-0.10}$} & \textgreater 1.00 \\ \cline{3-3} \cline{6-6} \cline{8-8} 
                         &                         & $6\times 10^5$ - $2\times 10^6$ &                         &                        & $-12.03^{+0.03}_{-0.03}$ &                       & \textgreater 1.12 \\ \cline{3-3} \cline{6-6} \cline{8-8} 
                         &                         & $2\times 10^6$ - $7\times 10^6$ &                         &                        & $-12.77^{+0.04}_{-0.05}$ &                       & \textgreater 1.17 \\ \cline{3-3} \cline{6-6} \cline{8-8} 
                         &                         & $10^7$ - $2\times 10^7$ &                         &                        & $-13.53^{+0.12}_{-0.14}$ &                       & \textgreater 0.68 \\ \hline
\multirow{3}{*}{130427A} & \multirow{3}{*}{0.34}   & $1\times 10^5$ - $7\times 10^5$ & \multirow{3}{*}{0.0164} & \multirow{3}{*}{0.073} & $-11.07^{+0.01}_{-0.01}$ & \multirow{3}{*}{$1.63^{+0.04}_{-0.05}$} & \textgreater 1.43 \\ \cline{3-3} \cline{6-6} \cline{8-8} 
                         &                         & $7\times 10^5$ - $4\times 10^6$ &                         &                        & $-12.04^{+0.02}_{-0.02}$ &                       & \textgreater 1.30 \\ \cline{3-3} \cline{6-6} \cline{8-8} 
                         &                         & $4\times 10^6$ - $2\times 10^7$ &                         &                        & $-12.99^{+0.05}_{-0.06}$ &                       & \textgreater 1.00 \\ \hline
\end{tabular}
\tablefoot{$^a$ Time intervals in which we analyze the spectral data. $^b$ Galactic  
hydrogen equivalent column density. $c$ Intrinsic hydrogen equivalent column density (at the source redshift). $^d$ Integrated flux in 0.3 - 10 keV energy band. $^e$ Photon 
spectral index. $^f$ One-side lower limit on $E_{\rm cut}$ at 2$\sigma$ confidence level. To construct  Fig.~\ref{fig:data}, we adopt, for each GRB,  the start time of the latest time interval and the corresponding 2$\sigma$ lower limit on $E_{\rm cut}$, thus deriving the most conservative constraints on the parameter space.}
\end{table*}

\section{Derivation of the dependency of the $h\nu_\mathrm{max}$ evolution on the jet break time $t_\mathrm{j}$}
\label{appendix:B}

In this Appendix we explicitly show  that the evolution of the maximum synchrotron frequency normalized to its value at the jet break, that is $h\nu_\mathrm{max}(t_\mathrm{obs})/h\nu_\mathrm{max}(t_\mathrm{j})$, depends only on $t_\mathrm{obs}/t_\mathrm{j}$. To see this, let us first demonstrate that, when $t_\mathrm{obs}>t_\mathrm{j}$, $h\nu_\mathrm{max}(t_\mathrm{obs})/h\nu_\mathrm{max}(t_\mathrm{j})$ depends only on $\Gamma\ \theta_\mathrm{j}$. In general, at a given latitude in the EATS, we can write
\begin{equation}
    h\nu_{\rm max}(t_\mathrm{obs},\theta) = \mathcal{D}(t_\mathrm{obs},\theta)\  \gamma_{\rm max}^{\prime 2}(R(t_\mathrm{obs},\theta))\frac{he \ B(R(t_\mathrm{obs},\theta))}{2\pi\  m_{\rm e}c}.
\end{equation}
It is straightforward to show that $\gamma_\mathrm{max}(R)\propto \Gamma(R)$ and that the magnetic field scales as $B(R)\propto \Gamma(R)$. Moreover, by using the small-angle approximation for $\cos\theta$ and the second-order expansion of $\beta$ in terms of $\Gamma$ in the definition of the Doppler factor, one finds $\mathcal{D}\sim 2\Gamma/(1+(\Gamma\theta)^2)$. For $t_\mathrm{obs}>t_\mathrm{j}$, the latitude that dominates the observed emission is $\theta=\theta_\mathrm{j}$, hence
\begin{equation}
    \frac{h\nu_\mathrm{max}(t_\mathrm{obs})}{h\nu_\mathrm{max}(t_\mathrm{j})}=\left(\frac{\Gamma(t_\mathrm{obs},\theta_\mathrm{j})}{\Gamma(t_\mathrm{j},\theta_\mathrm{j})}\right)^4\frac{1+\Gamma^2(t_\mathrm{j}, \theta_\mathrm{j})\ \theta_{\rm j}^2}{1+\Gamma^2(t_\mathrm{obs},\theta_\mathrm{j})\ \theta_{\rm j}^2}. 
\end{equation}
Because $\Gamma(t_\mathrm{j},\theta_\mathrm{j})=\theta_\mathrm{j}^{-1}$, the above expression becomes 
\begin{equation}
    \frac{h\nu_\mathrm{max}(t_\mathrm{obs})}{h\nu_\mathrm{max}(t_\mathrm{j})}=2\frac{\left(\Gamma(t_\mathrm{obs},\theta_\mathrm{j})\ \theta_\mathrm{j}\right)^4}{1+\Gamma^2(t_\mathrm{obs},\theta_\mathrm{j})\ \theta_{\rm j}^2},
\end{equation}
which is a function of $\Gamma\theta_\mathrm{j}$ only.

We now demonstrate that, in turn, $\Gamma\theta_\mathrm{j}$ depends only on $t_\mathrm{obs}/t_\mathrm{j}$. First, we
combine Eqs. \ref{eq:Gsh_of_R} and \ref{eq:r} to write a proportionality relation between $\Gamma_\mathrm{sh}$, $\theta$ and $t_\mathrm{obs}$ on the EATS, that is 
\begin{equation}
    t_\mathrm{obs} \propto \left[\frac{E_\mathrm{k}}{A}\right]^{1/(3-k)}\Gamma_\mathrm{sh}^{-2/(3-k)}(\theta)\left[1-\cos\theta+\frac{1}{2(4-k)\ \Gamma_\mathrm{sh}^2(\theta)}\right].
\end{equation}
Let us take the above relation evaluated at $\theta_\mathrm{j}$, divide both sides by $t_\mathrm{j}\propto (E_\mathrm{k}/A)^{1/(3-k)}\ \theta_\mathrm{j}^{2(4-k)/(3-k)}$, and use the small-angle approximation $\cos\theta_\mathrm{j}\approx 1-\theta_\mathrm{j}^2/2$. After some algebraic manipulation, we obtain
\begin{equation}
    \frac{t_\mathrm{obs}}{t_\mathrm{j}}\propto \left(\Gamma_\mathrm{sh}\theta_\mathrm{j}\right)^{-2(4-k)/(3-k)}\left[1+\left(\Gamma_\mathrm{sh}\theta_\mathrm{j}\right)^2\right]\equiv \mathcal{F}(\Gamma\theta_\mathrm{j}),
\end{equation}
where we used the fact that $\Gamma_\mathrm{sh}\sim \sqrt{2}\Gamma$ to express the right-hand side in terms of the quantity $\Gamma\theta_\mathrm{j}$.
Since the function $\mathcal{F}(x)$ is monotonic, it can be inverted, so that $\Gamma\theta_\mathrm{j} \propto \mathcal{F}^{-1}(t_\mathrm{obs}/t_\mathrm{j})$. This result demonstrates that the ratio $h\nu_\mathrm{max}(t_\mathrm{obs})/h\nu_\mathrm{max}(t_\mathrm{j})$ depends only on the ratio $t_\mathrm{obs}/t_\mathrm{j}$.

\end{appendix}

\end{document}